\documentclass[12pt,a4paper]{JHEP3}

\def\p{\partial}

\def\half{{1\over 2}}

\def\ah{{\hat\alpha}}

\def\bh{{\hat\beta}}
\def\gh{{\hat\gamma}}

\def\r{{\rho}}

\def\a{\alpha}
\def\b{\beta}
\def\g{\gamma}
\def\d{\delta}
\def\e{\epsilon}
\def\k{\kappa}
\def\m{\mu}

\def\n{\nu}
\def\l{\lambda}

\def\r{\rho}
\def\s{\sigma}
\def\p{\partial}
\def\t{{\theta}}

\def\S{\Sigma}

\def\O{\Omega}

\def\ah{{\hat \alpha}}

\newcommand{\be}{\begin{equation}}
\newcommand{\ee}{\end{equation}}
\newcommand{\bea}{\begin{eqnarray}}
\newcommand{\eea}{\end{eqnarray}}
\newcommand{\Str}{\ensuremath{\mathrm{Str}}}

\title{On Integrable Backgrounds Self-dual under Fermionic T-duality}

\author{Ido~Adam$^1$, Amit~Dekel$^2$ and Yaron~Oz$^2$\\
$^1$ Max-Planck-Institut f\"ur Gravitationsphysik\\
Albert Einstein Institut\\
Am M\"ulehnberg 1, 14476 Golm, Germany\\
and\\
Kavli Institute for Theoretical Physics\\
University of California\\
Santa Barbara, CA 93106-4030, USA\\
$^2$Raymond and Beverly Sackler School of Physics and Astronomy \\
Tel-Aviv University, Ramat-Aviv 69978, Israel \\
E-mails: \email{idoadam@aei.mpg.de}, \email{amitde@post.tau.ac.il},
\email{yaronoz@post.tau.ac.il}}

\abstract{We study the fermionic T-duality symmetry of integrable
  Green-Schwarz sigma-models on AdS backgrounds with Ramond-Ramond
  fluxes in various dimensions. We show that sigma-models based on
  supercosets of $\mathrm{PSU}$ supergroups, such as $AdS_2 \times
  S^2$ and $AdS_3\times S^3$ are self-dual under fermionic T-duality,
  while supercosets of OSp supergroups such as non-critical $AdS_2$
  and $AdS_4$ models, and the critical $AdS_4 \times \mathbb{C}P^3$
  background are not.  We present a general algebraic argument to when
  a supercoset is expected to have a fermionic T-duality symmetry, and
  when it will fail to have one.  }


\preprint{AEI-2009-020\\NSF-KITP-09-17}

\begin{document}

\section{Introduction and summary}

Recently, Alday and Maldacena proposed \cite{Alday:2007hr} that
planar $\mathcal{N}=4$ SU(N) SYM MHV gluon scattering
amplitudes at leading order in the strong 't~Hooft coupling expansion can be calculated using the dual
gravity (string) description. A crucial step in the calculation procedure is an application of an ordinary bosonic T-duality transformation to the four CFT
coordinates of $AdS_5 \times S^5$. This suggestion implies that such
amplitudes possess a dual conformal symmetry at strong coupling
originating from the fact that $AdS_5$ is self-dual under the
T-duality. This dual symmetry has been observed also in gluon scattering
amplitudes calculated in the weakly-coupled gauge theory description
\cite{Drummond:2006rz,Drummond:2007aua}.

Using the AdS/CFT duality it was shown that such a symmetry is
expected to be valid at all values of the 't~Hooft coupling
\cite{Berkovits:2008ic,Beisert:2008iq}. This was done by proving that
the aforementioned T-duality together with a novel T-duality of
Grassmann-odd coordinates of the target-superspace form an exact
quantum duality under which the full $AdS_5 \times S^5$ superstring is
self-dual. Since both the original background and its dual possess superconformal symmetry, it means that each of them
has both the manifest superconformal symmetry and a dual
one.
Furthermore, \cite{Berkovits:2008ic,Beisert:2008iq} have linked the
duals of the superconformal Noether currents to the non-local currents
implied by the integrability of the superstring on $AdS_5 \times S^5$
\cite{Bena:2003wd}.

In view of these results it is natural to inquire how ubiquitous this
dual superconformal symmetry is. The aim of this paper is to consider
this question by analyzing the fermionic T-duality symmetry of
integrable Green-Schwarz sigma-models on AdS backgrounds with Ramond-Ramond fluxes
in various dimensions. Some of these sigma-models have been constructed in \cite{Adam:2007ws,Zhou:1999sm,Park:1998un,Rahmfeld:1998zn}.

We show that sigma-models based on supercosets of $\mathrm{PSU}$
supergroups, such as $AdS_2 \times S^2$ and $AdS_3\times S^3$ are
self-dual under fermionic T-duality.  Supercosets of OSp supergroups
such as non-critical $AdS_2$ and $AdS_4$ models, and the critical
$AdS_4 \times \mathbb{C}P^3$ background (whose coset model was
constructed and explored in
\cite{Arutyunov:2008if,Stefanski:2008ik,Fre:2008qc,Bonelli:2008us,DAuria:2008cw})
are not self-dual.  In the OSp models we find that the Buscher
procedure \cite{Buscher:1987sk,Buscher:1987qj} fails due to a lack of
appropriate quadratic terms. This is because the Cartan-Killing
bilinear form of the ortho-symplectic group is non-zero only for
products of different Grassmann-odd generators. Thus, one may expect
this to imply that in those cases in which a dual theory exists, the
theory does not have a dual superconformal symmetry.

The paper is organized as follows.
In section 2
we show that the $AdS_p \times S^p$ ($p=2, 3$)
target-spaces based on PSU cosets are self-dual under a combination of
bosonic and fermionic T-duality.
In section 3 we consider models based on supercosets of the
ortho-symplectic supergroup, for which the Buscher procedure
\cite{Buscher:1987sk,Buscher:1987qj} of gauging an isometry of the
target-space in order to obtain the T-dual sigma-model fails.
These include the non-critical superstring on
$AdS_2$ with four supersymmetries and $AdS_4$ with eight supersymmetries, the supercoset construction
of $AdS_4 \times \mathbb{C}\mathrm{P}^3$, and a model of $AdS_2$ with eight supersymmetries.
In section 4 we present a general algebraic argument to when a supercoset is expected to have a
fermionic T-duality symmetry, and when it will fail to have one.
In the appendices we provide details on the relevant superalgebras that are used in the paper.

\section{$AdS_p \times S^p$ target-spaces}

In this section we show that the $AdS_p \times S^p$ ($p=2, 3$)
target-spaces based on PSU supercosets are self-dual under a combination of
bosonic and fermionic T-duality.

\subsection{The $AdS_2 \times S^2$
  target-space} \label{sec:AdS_2xS^2_target_space}

The target superspace whose bosonic part is $AdS_2 \times S^2$ can be
realized as the coset space $\mathrm{PSU}(1,1|2) / (\mathrm{U}(1) \times
\mathrm{U}(1))$\footnote{For this to be a superstring, this superspace has to be
supplemented by an additional internal CFT with the appropriate
central charge. It is not clear that such a description exists
using such a partial Green-Schwarz action (the hybrid model discussed
in \cite{Berkovits:1999zq} might be a better option for this kind of
superstring construction).}.
The Green-Schwarz sigma-model for supercosets with a $\mathbb{Z}_4 $
automorphism is given by the action \cite{Adam:2007ws}
\begin{equation} \label{eq:GS-coset-action}
  S = \frac{R^2}{4 \pi \alpha'} \int d^2 z \Str \left( J_2 \bar J_2 +
  \frac{1}{2} J_1 \bar J_3 - \frac{1}{2} \bar J_1 J_3 \right) \ ,
\end{equation}
where $J = g^{-1} \partial g$ for $g \in G$ and $J_i$ is the current
$J$ restricted to the invariant subspace $\mathcal{H}_i$ of the
$\mathbb{Z}_4$ automorphism of the algebra of the group $G$. Using the
$\mathrm{psu(1,1|2)}$ algebra given in appendix
\ref{sec:psu112-algebra}, the sigma-model (\ref{eq:GS-coset-action})
takes the form
\begin{eqnarray}
  S & = & \frac{R^2}{4 \pi \alpha'} \int d^2 z \Big[ \frac{1}{2} (J_P
    - J_K) (\bar J_P - \bar J_K) + \frac{1}{2} J_D \bar J_D +
    \frac{1}{2} J_{R_1} \bar J_{R_1} + \frac{1}{2} J_{R_2} \bar
    J_{R_2} - \nonumber \\
    && {} - \frac{i}{2} \eta_{\alpha \beta} (J_{Q_\alpha} \bar
    J_{Q_\beta} - J_{\hat Q_\alpha} \bar J_{\hat Q_\beta} +
    J_{S_\alpha} \bar J_{S_\beta} - J_{\hat S_\alpha} \bar J_{\hat
      S_\beta}) \Big] \ ,
\end{eqnarray}
where $\eta_{1 2} = \eta_{2 1} = 1$ and zero otherwise. The analysis
here will follow that of \cite{Berkovits:2008ic}. A general group
element $g \in \mathrm{PSU(1, 1|2)}$ can be parameterized as
\begin{equation} \label{eq:psu112-group-element}
  g = e^{x P + x' K + \theta^\alpha Q_\alpha + \xi^\alpha S_\alpha}
  e^B \ , \quad
  e^B \equiv e^{\hat \theta^\alpha \hat Q_\alpha + \hat \xi^\alpha \hat
  S_\alpha} y^D e^{\sum y^i R_i / y} \ .
\end{equation}
We partially fix the $\kappa$-symmetry such that $\xi^\alpha = 0$, and
we use the $\mathrm{U}(1) \times \mathrm{U}(1)$ gauge symmetry
generated by $P + K$ and $R_3$ to set $x' = 0$, thus the coset
representative is
\begin{equation}
  g = e^{x P + \theta^\alpha Q_\alpha} e^B \ .
\end{equation}
Using the fact that the group generated by $\{ \hat Q, \hat S, D, R_i
\}$ transforms $P$ and $Q_\alpha$ among themselves, the components of
the Maurer-Cartan 1-form are
\begin{eqnarray}
  J_P & = & [e^{-B} (dx P + d\theta^\alpha Q_\alpha) e^B]_P \ , \quad
  J_{Q_\alpha} = [e^{-B} (dx P + d\theta^\beta Q_\beta)
    e^B]_{Q_\alpha} \ , \nonumber \\
  J_K & = & 0 \ , \quad
  J_{\hat Q_\alpha} = [e^{-B} de^B]_{\hat Q_\alpha} \ , \quad
  J_{S_\alpha} = 0 \ , \quad
  J_{\hat S_\alpha} = [e^{-B} de^B]_{\hat S_\alpha} \ , \nonumber \\
  J_D & = & [e^{-B} de^B]_D \ , \quad
  J_{R_i} = [e^{-B} de^B]_{R_i} \ .
  \label{eq:psu112-Maurer-Cartan-1-form-components}
\end{eqnarray}

This sigma-model is T-dualized in the directions of the Abelian
sub-algebra formed by the generators $P$ and $Q_\alpha$ according to
the procedure of \cite{Buscher:1987sk,Buscher:1987qj}, by introducing
the gauge fields $A$, $\bar A$ for the translation $P$ and $A^\alpha$,
$\bar A^\alpha$ for the supercharges $Q_\alpha$ and the corresponding
Lagrange multipliers $\tilde x$ and $\tilde \theta_\alpha$:
\begin{eqnarray}
  S & = & \frac{R^2}{4 \pi \alpha'} \int d^2 z \Big[ \frac{1}{2}
    [e^{-B} (A P + A^\alpha Q_\alpha) e^B]_P [e^{-B} (\bar A P + \bar
      A^\alpha Q_\alpha) e^B]_P - \nonumber \\
    && {} -\frac{i}{2} \eta_{\alpha \beta}
    [e^{-B} (A P + A^\gamma Q_\gamma) e^B]_{Q_\alpha} [e^{-B} (\bar A
      P + \bar A^\gamma Q_\gamma) e^B]_{Q_\beta} + \nonumber \\
    && {} + \frac{1}{2} J_D
    \bar J_D + \frac{1}{2} J_{R_1} \bar J_{R_1} + \frac{1}{2} J_{R_2}
    \bar J_{R_2} + \frac{i}{2} \eta_{\alpha \beta} (J_{\hat Q_\alpha}
      \bar J_{\hat Q_\beta} + J_{\hat S_\alpha} \bar J_{\hat S_\beta})
      + \nonumber \\
      && {} + \tilde x (\bar \partial A - \partial \bar A) + \tilde
      \theta_\alpha (\bar \partial A^\alpha - \partial \bar A^\alpha)
      \Big] \ .
\end{eqnarray}
It is convenient to change variables to
\begin{equation} \label{eq:ps112-new-variables}
  A' = [e^{-B} (A P + A^\alpha Q_\alpha) e^B ]_P \ , \quad
  A'^\alpha = [e^{-B} (A P + A^\beta Q_\beta) e^B]_{Q_\alpha}
\end{equation}
and similarly for the right-moving gauge fields. Using the inverted
relations
\begin{equation}
  A = [e^B (A' P + A'^\alpha Q_\alpha) e^{-B}]_P \ , \quad
  A^\alpha = [e^B (A' P + A'^\beta Q_\beta) e^{-B}]_{Q_\alpha} \ ,
\end{equation}
the action in terms of the new variables reads
\begin{eqnarray}
  S & = & \frac{R^2}{4 \pi \alpha'} \int d^2 z \Big[ \frac{1}{2} A'
    \bar A' - \frac{i}{2} \eta_{\alpha \beta} A'^\alpha \bar A'^\beta
    + \dots - \nonumber \\
    && {} - \bar \partial \tilde x (A' [e^B P e^{-B}]_P + A'^\alpha [e^B
      Q_\alpha e^{-B}]_P) + \nonumber \\
    && {} + \partial \tilde x \left( \bar A' [e^B P e^{-B}]_P + \bar
    A'^\alpha [e^B Q_\alpha e^{-B}]_P \right) - \nonumber \\
    && {} - \bar \partial \tilde \theta_\alpha \left( A' [e^B P
      e^{-B}]_{Q_\alpha} + A'^\beta [e^B Q_\beta e^{-B}]_{Q_\alpha}
    \right) + \nonumber \\
    && {} + \partial \tilde \theta_\alpha \left( \bar A' [e^B P
      e^{-B}]_{Q_\alpha} + \bar A'^\beta [e^B Q_\beta
      e^{-B}]_{Q_\alpha} \right) \Big] \ ,
\end{eqnarray}
where $\dots$ denotes the spectator terms. Since the gauge fields
appear quadratically in the action, one can integrate them out by
substituting their equations of motion
\begin{eqnarray}
  A' & = & -2 [e^B \partial \tilde x P e^{-B}]_P - 2[e^B \partial
    \tilde \theta_\alpha P e^{-B}]_{Q_\alpha} = -2 [e^{-B} (\partial
    \tilde x K + i \partial \tilde \theta_\alpha S^\alpha) e^B]_K \ ,
  \nonumber \\
  \bar A' & = & 2 [e^B \bar \partial \tilde x P e^{-B}]_P + 2 [e^B
    \bar \partial \tilde \theta_\alpha P e^{-B}]_{Q_\alpha} =
  2 [e^{-B} (\bar \partial \tilde x K + i \bar \partial \tilde
    \theta_\alpha S^\alpha) e^B]_K \ , \nonumber \\
  A'^\alpha & = & 2 i \eta^{\alpha \beta} ([e^B \partial \tilde x
    Q_\beta e^{-B}]_P - [e^B \partial \tilde \theta_\gamma Q_\beta
    e^{-B}]_{Q_\gamma}) =
  -2 \eta^{\alpha \beta} [e^{-B} (\partial \tilde x K + i \partial
    \tilde \theta_\gamma S^\gamma) e^B]_{S^\beta}  , \nonumber \\
  \bar A'^\alpha & = & 2 i \eta^{\alpha \beta} ([e^B \bar \partial
    \tilde x Q_\beta e^{-B}]_P - [e^B \bar \partial \tilde
    \theta_\gamma Q_\beta e^{-B}]_{Q_\gamma}) =
  -2 \eta^{\alpha \beta} [e^{-B} (\bar \partial \tilde x K + i \bar
    \partial \tilde \theta_\gamma S^\gamma) e^B]_{S^\beta}
    . \nonumber \\
  {}
\end{eqnarray}
 and obtain after rescaling $\tilde x \to \frac{1}{2}
\tilde x$ and $\tilde \theta_\alpha \to \frac{1}{2} \tilde \theta_\alpha$
\begin{eqnarray}
  S & = & \frac{R^2}{4 \pi \alpha'} \int d^2 z \Big[ \frac{1}{2}[e^{-B}
      (\partial \tilde x K + i \partial \tilde \theta_\alpha S^\alpha)
      e^B]_K [e^{-B} (\bar \partial \tilde x K + i \bar \partial
      \tilde \theta_\alpha S^\alpha) e^B]_K - \nonumber \\
    && {} - \frac{i}{2} \eta_{\alpha \beta} [e^{-B} (\partial \tilde x
      K + i \partial \tilde \theta_\gamma S^\gamma) e^B]_{S_\alpha}
    [e^{-B} (\bar \partial \tilde x K + i \bar \partial \tilde
      \theta_\gamma S^\gamma) e^B]_{S_\beta} + \dots \Big] \ ,
\label{eq:T-dualized-psu112-action}
\end{eqnarray}
where we used $\epsilon_{\alpha \beta}$ to lower the spinor indices
of $\eta^{\alpha \beta}$.

In order to show the self-duality of this background under the above
T-duality, the original action has to be brought to the same form as
(\ref{eq:T-dualized-psu112-action}). One can easily check by using the
$\epsilon_{\beta \gamma} \sigma^{i j \gamma}_\alpha =
\epsilon_{\alpha \gamma} \sigma^{i j \gamma}_\beta$ that the
$\mathrm{psu}(1, 1|2)$ algebra admits the automorphism
\begin{equation}
  P \leftrightarrow K \ ,
  \quad D \to -D \ ,
  \quad  Q_\alpha \leftrightarrow S^\alpha \ , \quad
  \hat Q_\alpha \leftrightarrow \hat S_\alpha \ ,
\end{equation}
with the rest of the generators unchanged. Applying this automorphism
combined with the change of variables
\begin{equation}
  x \to \tilde x \ , \quad
  \theta^\alpha \to i \tilde \theta_\alpha \ , \quad
  \hat \theta^\alpha \leftrightarrow \hat \xi^\alpha \ , \quad
  y_i \to \frac{y_i}{y^2} \ ,
\end{equation}
to (\ref{eq:psu112-Maurer-Cartan-1-form-components}) one obtains
(\ref{eq:T-dualized-psu112-action}).

In order to complete the proof of quantum mechanical equivalence we
also have to show that the Jacobian functional determinant from the
change of variables (\ref{eq:ps112-new-variables}) is the
identity. The transformation of variables was done using $e^B$. Since
it is in a unitary subgroup of the $\mathrm{PSU}(1,1|2)$ group, its
super-determinant is equal to one and hence the Jacobian of the
transformation is trivial and does not induce any shift of the
dilaton. In order to see this explicitly, one can treat $(A,
A^\alpha)$ as a vector and write the adjoint action of the generators
$D$, $R_{i j}$, $\hat Q_\alpha$ and $\hat S_\alpha$ as matrices acting
on this vector
\begin{eqnarray}
  [\mathrm{ad}_D] & = & \left(
  \begin{array}{cc}
    1 & 0 \\
    0 & \frac{1}{2} {\delta_\alpha}^\beta
  \end{array}
  \right) \ , \quad
  [\mathrm{ad}_{R_{i j}}] = \left(
    \begin{array}{cc}
      0 & 0 \\
      0 & -\frac{1}{2} {\sigma_{i j \alpha}}^\beta
    \end{array}
    \right) \  , \nonumber \\
  {}[\mathrm{ad}_{\hat Q_\alpha}] & = & \left(
  \begin{array}{cc}
    0 & -\epsilon_{\alpha \beta} \\
    0 & 0
  \end{array}
  \right) \ , \quad
  [\mathrm{ad}_{\hat S_\alpha}] = \left(
  \begin{array}{cc}
    0 & 0 \\
    -i {\delta_\alpha}^\beta & 0
  \end{array}
  \right) \ ,
\end{eqnarray}
which evidently have a vanishing supertrace. Hence, the
super-determinant of similarity transformations with elements of the
group generated by these generators is the identity.

\subsection{The $AdS_3 \times S^3$ target-space}
We construct the Green-Schwarz sigma-model on $AdS_3\times S^3$ using the supercoset manifold
$(\mathrm{PSU}(1,1|2) \times \mathrm{PSU}(1,1|2))
/(\mathrm{SU}(1,1)\times \mathrm{SU}(2))$ with 16 supersymmetry
generators. Using the $\mathbb{Z}_4$ structure
(\ref{eq:ads3S3Z4struct}) we have \be J_2=J^-_{a}(P_a-K_a)+J_D D+J_\m
R_\m =\frac{1}{2}(J_{P_a}-J_{K_a})(P_a-K_a)+J_D D+J_\m R_\m \ee
$$
J_1=J_{S^{I}_{\a\ah}+a^{IJ}Q^{J}_{\a\ah}}(S^{I}_{\a\ah}+a^{IK}Q^{K}_{\a\ah})
=\frac{1}{2}(J_{S^{I}_{\a\ah}}+a^{IJ}J_{Q^{J}_{\a\ah}})(S^{I}_{\a\ah}+a^{IK}Q^{K}_{\a\ah})
$$
$$
J_3=J_{S^{I}_{\a\ah}-a^{IJ}Q^{J}_{\a\ah}}(S^{I}_{\a\ah}-a^{IK}Q^{K}_{\a\ah})
=\frac{1}{2}(J_{S^{I}_{\a\ah}}-a^{IJ}J_{Q^{J}_{\a\ah}})(S^{I}_{\a\ah}-a^{IK}Q^{K}_{\a\ah}) \ .
$$
Thus,
\be
\Str(J_2\bar J_2)=-\frac{1}{2}(J_{P_a}-J_{K_a})(\bar J_{P_b}-\bar J_{K_b})\eta^{ab}
+J_D\bar J_D+J_{R_\m}\bar J_{R_\m} \ ,
\ee
\be
\Str(J_1\bar J_3)=\frac{1}{2}(\s^3)^{IK}\e_{\a\b}\e_{\ah\bh}
(J_{S^{I}_{\a\ah}}+a^{IJ}J_{Q^{J}_{\a\ah}})
(\bar J_{S^{K}_{\b\bh}}-a^{KL}\bar J_{Q^{L}_{\b\bh}}) \ ,
\ee
and the action reads
\be
S=\frac{R^2}{4\pi\a'}\int d^2z
\bigg[-\frac{1}{2}(J_{P_a}-J_{K_a})(\bar J_{P_b}-\bar J_{K_b})\eta_{ab}
+J_D\bar J_D+J_{R_\m}\bar J_{R_\m}
\ee
$$
+\frac{1}{2}(\s^3)^{IK}\e_{\ah\bh}\e_{\a\b}
\bigg(J_{S^{I}_{\a\ah}}\bar J_{S^{K}_{\b\bh}}
+J_{Q^{I}_{\a\ah}}\bar J_{Q^{K}_{\b\bh}}\bigg)\bigg] \ .
$$
Next we parameterize the supergroup element
\be
g=\exp(x^a P_a + \t^{\a\ah 1}Q_{\a\ah}^{ 1})\exp(\t^{\a\ah 2}Q_{\a\ah}^{ 2} + \xi^{\a\ah 1}S_{\a\ah}^{ 1})y^D\exp(y_\m R_\m/y)
\ee
$$
\equiv\exp(x^a P_a + \t^{\a\ah 1}Q_{\a\ah}^{ 1})e^B \ .
$$
Define the currents
\be\label{current_separation}
J=g^{-1}dg=j+\mathfrak{j}, \qquad
j=e^{-B}(dx^a P_a + d\t^{\a\ah 1}Q_{\a\ah}^{ 1})e^B, \qquad
\mathfrak{j}=e^{-B} d e^B \ .
\ee
Using these definitions and the algebra (\ref{eq:psu(1,1|2)^2Alg}) we get the currents
\be
J_{P_a}=j_{P_a},\qquad
J_{Q_{\a\ah}^{ 1}}=j_{Q_{\a\ah}^{ 1}},\qquad
J_{Q_{\a\ah}^{ 2}}=\mathfrak{j}_{Q_{\a\ah}^{ 2}},\qquad
J_{S_{\a\ah}^{ 1}}=\mathfrak{j}_{S_{\a\ah}^{ 1}},\qquad
J_{S_{\a\ah}^{ 2}}=0
\ee
$$
J_{K_a}=0,\qquad
J_{D}=\mathfrak{j}_D,\qquad
J_{R_\m}=\mathfrak{j}_{R_\m},\qquad
$$
the action reads
\be\label{eq:Ads3S3action}
S=\frac{R^2}{4\pi\a'}\int d^2z
\bigg[-\frac{1}{2}j_{P_a}\bar j_{P_b}\eta_{ab}
+\mathfrak{j}_D\bar \mathfrak{j}_D+\mathfrak{j}_{R_\m}\bar \mathfrak{j}_{R_\m}
\ee
$$
+\frac{1}{2}\e_{\a\b}\e_{\ah\bh}
\bigg(\mathfrak{j}_{S^{1}_{\a\ah}}\bar \mathfrak{j}_{S^{1}_{\b\bh}}
+j_{Q^{1}_{\a\ah}}\bar j_{Q^{1}_{\b\bh}}
-\mathfrak{j}_{Q^{2}_{\a\ah}}\bar \mathfrak{j}_{Q^{2}_{\b\bh}}
\bigg)
\bigg] \ .
$$
Introducing gauge fields and Lagrange multipliers we get the action
\be
S=\frac{R^2}{4\pi\a'}\int d^2z
\bigg[-\frac{1}{2}A'_{P_a}\bar A'_{P_b}\eta_{ab}
+\mathfrak{j}_D\bar \mathfrak{j}_D+\mathfrak{j}_{R_\m}\bar \mathfrak{j}_{R_\m}
\ee
$$
+\frac{1}{2}\e_{\a\b}\e_{\ah\bh}
\bigg(\mathfrak{j}_{S^{1}_{\a\ah}}\bar \mathfrak{j}_{S^{1}_{\b\bh}}
+A'_{Q^{1}_{\a\ah}}\bar A'_{Q^{1}_{\b\bh}}
-\mathfrak{j}_{Q^{2}_{\a\ah}}\bar \mathfrak{j}_{Q^{2}_{\b\bh}}
\bigg)
$$
$$
+\tilde x_a(\bar\p A^a-\p\bar A^a)
+\tilde \t^{1}_{\a\ah}(\bar\p A_{Q^{1}_{\a\ah}}-\p\bar A_{Q^{1}_{\a\ah}})
\bigg] \ ,
$$
where
\be
A'=e^{-B}(A^a P_a + A^{1\a\ah}Q^{1}_{\a\ah})e^B,\qquad
A^{1\a\ah} \equiv A_{Q^{1}_{\a\ah}},\qquad
A_{P_a} \equiv A^a,\qquad
\ee
$$
A=e^{B}(A'^a P_a + A'^{1\a\ah}Q^{1}_{\a\ah})e^{-B} \ .
$$
Solving for $A'$ we get
\be
A'_a=2(\p\tilde x_b[e^{B}P_ae^{-B}]_{P_b}+\p\tilde \t^{1}_{\a\ah}[e^{B}P_ae^{-B}]_{Q^{1}_{\a\ah}})
\ee
$$
=2\eta_{ac}[e^{-B}(\p\tilde x^b K_b+\p\tilde \t^{1\a\ah}S^2_{\a\ah})e^{B}]_{K_c}
$$
\be
\bar A'_a=-2(\bar \p\tilde x_b[e^{B}P_ae^{-B}]_{P_b}+\bar \p\tilde \t^{1}_{\a\ah}[e^{B}P_ae^{-B}]_{Q^{1}_{\a\ah}})
\ee
$$
=-2\eta_{ac}[e^{-B}(\bar\p\tilde x^b K_b+\bar\p\tilde \t^{1\a\ah}S^2_{\a\ah})e^{B}]_{K_c}
$$
\be
\frac{1}{2}\e_{\a\b}\e_{\ah\bh}A'_{Q^{1}_{\b\bh}}=-(\p\tilde x_b[e^{B}Q^{1}_{\a\ah} e^{-B}]_{P_b}-\p\tilde \t^{1}_{\b\bh}[e^{B}Q^{1}_{\a\ah} e^{-B}]_{Q^{1}_{\b\bh}})
\ee
$$
=-\e_{\a\b}\e_{\ah\bh}[e^{-B}(\p\tilde x^b K_b+\p\tilde \t^{1\g\gh}S^2_{\g\gh})e^{B}]_{S^2_{\b\bh}}
$$
\be
\frac{1}{2}\e_{\a\b}\e_{\ah\bh}\bar A'_{Q^{1}_{\b\bh}}=-(\bar \p\tilde x_b[e^{B}Q^{1}_{\a\ah} e^{-B}]_{P_b}-\bar \p\tilde \t^{1}_{\b\bh}[e^{B}Q^{1}_{\a\ah} e^{-B}]_{Q^{1}_{\b\bh}})
\ee
$$
=-\e_{\a\b}\e_{\ah\bh}[e^{-B}(\bar\p\tilde x^b K_b+\bar\p\tilde \t^{1\g\gh}S^2_{\g\gh})e^{B}]_{S^2_{\b\bh}} \ .
$$
Thus,
\be
S=\frac{R^2}{4\pi\a'}\int d^2z
\bigg[\frac{1}{2}A'_{P_a}\bar A'_{P_b}\eta_{ab}
+\mathfrak{j}_D\bar \mathfrak{j}_D+\mathfrak{j}_{R_\m}\bar \mathfrak{j}_{R_\m}
\ee
$$
+\frac{1}{2}\e_{\a\b}\e_{\ah\bh}
\bigg(\mathfrak{j}_{S^{1}_{\a\ah}}\bar \mathfrak{j}_{S^{1}_{\b\bh}}
-A'_{Q^{1}_{\a\ah}}\bar A'_{Q^{1}_{\b\bh}}
-\mathfrak{j}_{Q^{2}_{\a\ah}}\bar \mathfrak{j}_{Q^{2}_{\b\bh}}
\bigg)
\bigg]
$$
$$
=\frac{R^2}{4\pi\a'}\int d^2z
\bigg[-\frac{1}{2}
4[e^{-B}(\p\tilde x^b K_b+\p\tilde \t^{1\a\ah}S^2_{\a\ah})e^{B}]_{K_a}
[e^{-B}(\bar\p\tilde x^b K_b+\bar\p\tilde \t^{1\a\ah}S^2_{\a\ah})e^{B}]_{K_b}
\eta^{ab}
$$
$$
+\mathfrak{j}_D\bar \mathfrak{j}_D+\mathfrak{j}_{R_\m}\bar \mathfrak{j}_{R_\m}
+\frac{1}{2}\e_{\a\b}\e_{\ah\bh}
\bigg(\mathfrak{j}_{S^{1}_{\a\ah}}\bar \mathfrak{j}_{S^{1}_{\b\bh}}
-\mathfrak{j}_{Q^{2}_{\a\ah}}\bar \mathfrak{j}_{Q^{2}_{\b\bh}}
$$
$$
-4[e^{-B}(\p\tilde x^b K_b+\p\tilde \t^{1\g\gh}S^2_{\g\gh})e^{B}]_{S^2_{\a\ah}}
[e^{-B}(\bar\p\tilde x^b K_b+\bar\p\tilde \t^{1\g\gh}S^2_{\g\gh})e^{B}]_{S^2_{\b\bh}}
\bigg)
\bigg] \ .
$$
We can define $J'$'s
\be
J'=e^{-B}(2\p\tilde x^b K_b+2\p\tilde \t^{1\g\gh}S^2_{\g\gh})e^{B} \ ,
\ee
and the action reads
\be\label{eq:TdualizedAds3S3action}
S=\frac{R^2}{4\pi\a'}\int d^2z
\bigg[-\frac{1}{2}J'_{P_a}\bar J'_{P_b}\eta_{ab}
+\mathfrak{j}_D\bar \mathfrak{j}_D+\mathfrak{j}_{R_\m}\bar \mathfrak{j}_{R_\m}
\ee
$$
+\frac{1}{2}\e_{\a\b}\e_{\ah\bh}
\bigg(\mathfrak{j}_{S^{1}_{\a\ah}}\bar \mathfrak{j}_{S^{1}_{\b\bh}}
-J'_{S^{2}_{\a\ah}}\bar J'_{S^{2}_{\b\bh}}
-\mathfrak{j}_{Q^{2}_{\a\ah}}\bar \mathfrak{j}_{QS^{2}_{\b\bh}}
\bigg)\bigg] \ .
$$
Applying the following automorphism of the algebra
\be
D\rightarrow-D,\quad
K_a\leftrightarrow -\s^3_{ab}P_b,\quad
J_{ab}\rightarrow-J_{ab},\quad
\ee
$$
R_1\rightarrow R_1,\quad
R_2\rightarrow -R_2,\quad
R_3\rightarrow -R_3,\quad
$$
$$
N_{12}\rightarrow -N_{12},\quad
N_{23}\rightarrow  N_{23},\quad
N_{31}\rightarrow -N_{31},\quad
$$
$$
Q^I_{\a\ah}\leftrightarrow \s^1_{IJ}\s^1_{\ah\bh}S^J_{\a\bh} \ ,
$$
followed by the change of variables
\be
x_a\rightarrow -2\s^3_{ab}\tilde x_b,\quad
\t^1_{\a\ah}\rightarrow 2\s^1_{\ah\bh}\tilde\t^2_{\a\bh},\quad
\ee
$$
\t^2_{\a\ah}\leftrightarrow \s^1_{\ah\bh}\xi^1_{\a\bh},\quad
y_1\rightarrow y_1/y^2,\quad
y_2\rightarrow -y_2/y^2,\quad
y_3\rightarrow -y_3/y^2,
$$
the action (\ref{eq:TdualizedAds3S3action}) is mapped to the original one (\ref{eq:Ads3S3action}).


\section{Models based on the ortho-symplectic supergroup}
In this section we consider models based on supercosets of the
ortho-symplectic supergroup, for which the Buscher procedure
\cite{Buscher:1987sk,Buscher:1987qj} of gauging an isometry of the
target-space in order to obtain the T-dual sigma-model fails.
These include the non-critical superstring on
$AdS_2$ with four supersymmetries and $AdS_4$ with eight supersymmetries, the supercoset construction
of $AdS_4 \times \mathbb{C}\mathrm{P}^3$ which is conjectured to be
dual to superconformal Chern-Simons theory in three dimensions
\cite{Aharony:2008ug}, and a model of $AdS_2$ with eight supersymmetries.

\subsection{The Green-Schwarz sigma-model on $AdS_2$ with four
  supersymmetries} \label{sec:AdS2-sigma-model}

The non-critical $AdS_2$ background with RR-flux can be
realized as the supercoset $\mathrm{OSp}(2|2)/(\mathrm{SO}(1,1)
\times \mathrm{SO}(2))$. The Green-Schwarz action is of the form
(\ref{eq:GS-coset-action}) (see Appendix \ref{sec:osp22-algebra} for
the details of the algebra and the conventions that are used).

Fixing the $\mathrm{SO}(1,1) \times \mathrm{SO}(2)$ gauge symmetry and
$\kappa$-symmetry
one can choose the coset representative
\begin{equation}
  g = e^{x P + \theta Q} e^{\bar \theta \bar Q + \bar \xi \bar S} y^D
  \ .
\end{equation}
The generators $P$ and $Q$ form an Abelian subalgebra, which we will
attempt to T-dualize. In this parameterization the Maurer-Cartan
current $J$ takes the form
\begin{eqnarray}
  J & = & \frac{1}{y} \left( \partial x - 2 \partial \theta \bar \theta
  \right) P + \frac{1}{y^{1/2}} \partial \theta Q + \frac{1}{y^{1/2}}
  \left[ \left( 2 \partial \theta \bar \theta - \partial x \right) \bar
    \xi  + \partial \bar \theta \right] \bar Q +
  y^{1/2} \partial \bar \xi \bar S + \nonumber \\
  && {} + \left( \frac{\partial y}{y} - 2 \partial \theta \bar \xi
  \right) D + i \partial \theta \bar \xi R \ .
\end{eqnarray}

The sigma-model can be written explicitly as
\begin{eqnarray}
  S & = & \frac{R^2}{4 \pi \alpha'} \int d^2 z \Big[ -\frac{1}{2}
    (J_P + J_K) (\bar J_P + \bar J_K) + \frac{1}{2} J_D \bar J_D + J_Q
    \bar J_{\bar Q} +J_{\bar Q} \bar J_Q - \nonumber \\
    && {} - J_S \bar J_{\bar S} - J_{\bar S} \bar J_S \Big] \ .
\end{eqnarray}
The action in terms of the coordinates is then given by
\begin{eqnarray}
  S & = & \frac{R^2}{4 \pi \alpha'} \int d^2 z \Bigg[ \frac{-\partial x
      \bar \partial x + \partial y \bar \partial y}{2 y^2} -
    \frac{1}{y^2} \bar \theta ( \partial x \bar \partial \theta +
    \partial \theta \bar \partial x) + \frac{1}{y} \bar \xi (\partial y \bar
    \partial \theta + \partial \theta \bar \partial y) + \nonumber \\
    && {} + \frac{4}{y} \bar \theta \bar \xi \partial \theta \bar
    \partial \theta + \frac{1}{y} \bar \xi (\partial \theta \bar
    \partial x - \partial x \bar \partial \theta) + \frac{1}{y}
    (\partial \theta \bar \partial \bar \theta
    + \partial \bar \theta \bar \partial \theta) \Bigg] \ .
\end{eqnarray}
Note that the action is indeed quadratic in $\theta$ so naively one should be
able to T-dualize it along that coordinate.

An equivalent action can be written using two gauge fields
\cite{Buscher:1987sk,Buscher:1987qj}
\begin{eqnarray}
  S & = & \frac{R^2}{4 \pi \alpha'} \int d^2 z \Bigg[ \frac{-A_x
    \bar A_x + \partial y \bar \partial y}{2 y^2} -
    \frac{1}{y^2} \bar \theta ( A_x \bar A_\theta +
    A_\theta \bar A_x) + \frac{1}{y} \bar \xi (\partial y \bar
    A_\theta + A_\theta \bar \partial y) + \nonumber \\
    && {} + \frac{4}{y} \bar \theta \bar \xi A_\theta \bar
    A_\theta + \frac{1}{y} \bar \xi (A_\theta \bar
    A_x - A_x \bar A_\theta) + \frac{1}{y}
    (A_\theta \bar \partial \bar \theta + \partial \bar \theta \bar
    A_\theta) + \tilde x (\bar \partial A_x - \partial \bar A_x) +
    \nonumber \\
    && {} + \tilde \theta (\bar \partial A_\theta - \partial \bar
    A_\theta) \Bigg] \ .
\end{eqnarray}
The classical equations of motion for the gauge fields are
\begin{eqnarray}
  \frac{1}{2 y^2} A_x + \frac{1}{y^2} \bar \theta
  A_\theta - \frac{1}{y} \bar \xi A_\theta - \partial \tilde x & = & 0
  \ , \\
  \frac{1}{2 y^2} \bar A_x + \frac{1}{y^2} \bar \theta \bar A_\theta +
  \frac{1}{y} \bar \xi \bar A_\theta + \bar \partial \tilde x & = &
  0 \ , \\
  \frac{1}{y^2} \bar \theta A_x - \frac{1}{y} \bar \xi \partial y
  - \frac{4}{y} \bar \theta \bar \xi A_\theta + \frac{1}{y} \bar \xi
  A_x - \frac{1}{y} \partial \bar \theta - \partial \tilde \theta & =
  & 0 \ , \\
  \frac{1}{y^2} \bar \theta \bar A_x - \frac{1}{y} \bar \xi \bar \partial y
  + \frac{4}{y} \bar \theta \bar \xi \bar A_\theta - \frac{1}{y} \bar \xi
  \bar A_x + \frac{1}{y} \bar \partial \bar \theta + \bar \partial
  \tilde \theta & = & 0  \ .
\end{eqnarray}
Since Grassmann variables such as $\bar \theta$ and $\bar \xi$ have
non-trivial kernels, one cannot solve for $A_\theta$ and for $\bar
A_\theta$. Solving for $A_x$ and $\bar A_x$ one gets
\begin{eqnarray}
  A_x & = & -2 \bar \theta A_\theta + 2 y \bar \xi A_\theta + 2 y^2
  \partial \tilde x \ , \label{eq:osp22-Ax-solution} \\
  \bar A_x & = & -2 \bar \theta \bar A_\theta - 2 y \bar \xi \bar
  A_\theta - 2 y^2 \bar \partial \tilde x \ . \label{eq:osp22-Atheta-solution}
\end{eqnarray}

Substituting (\ref{eq:osp22-Ax-solution}) and
(\ref{eq:osp22-Atheta-solution}) in the action (this can be done since
they appear only quadratically in the action, but the classical
procedure should be supplemented by a functional determinant coming
from the Gaussian path integration of $A_x$ and $\bar A_x$) yields
\begin{eqnarray}
  S & = & \frac{R^2}{4 \pi \alpha'} \int d^2 z \Big[ -2 y^2 \partial
    \tilde x \bar \partial \tilde x - 2 y \bar \xi \partial \tilde x
    \bar A_\theta - 2y \bar \xi A_\theta \bar \partial \tilde x - 2
    \bar \theta \partial \tilde x \bar A_\theta + 2 \bar \theta
    A_\theta \bar \partial \tilde x  + \nonumber \\
    && {} + \frac{\partial y \bar \partial y}{2 y^2} + \frac{1}{y}
    \bar \xi (\partial y \bar A_\theta + A_\theta \bar \partial y) +
    \frac{1}{y} (A_\theta \bar \partial \bar \theta + \partial \bar
    \theta \bar A_\theta) - \bar \partial \tilde \theta A_\theta +
    \partial \tilde \theta \bar A_\theta \Big] \ .
\end{eqnarray}
Thus, after integrating out $A_x$ and $\bar A_x$ the remaining
fermionic gauge fields $A_\theta$ and $\bar A_\theta$ serve as
Lagrange multipliers forcing in the path integration over $\tilde
\theta$ that
\begin{eqnarray}
  2 \bar \theta \partial \tilde x - \frac{1}{y} \bar \xi \partial y +
  2 y \bar \xi \partial \tilde x  - \frac{1}{y} \partial \bar
  \theta - \partial \tilde \theta & = & 0 \ , \nonumber \\
  2 \bar \theta \bar \partial \tilde x + \frac{1}{y} \bar \xi \bar
  \partial y - 2 y \bar \xi \bar \partial \tilde x - \frac{1}{y} \bar
  \partial \bar \theta - \bar \partial \tilde \theta & = & 0\ ,
\end{eqnarray}
which express the non-zero modes of $\tilde \theta$ in terms of the
other fields, effectively reducing the path integration over $\tilde
\theta$ only to its zero-modes.

This appears rather strange as the original action did include a term
quadratic in $\theta$. A possible explanation is that the
quadratic term is $\kappa$-symmetry-exact, and hence does not influence
the equations of motion.

In order to support the above claim we do the same computation with a
different gauge choice for the $\kappa$-symmetry. The coset
representative in this gauge can be parameterized as
\begin{equation}
  g = e^{x P + \theta Q} e^{\bar \theta \bar Q + \xi S} y^D \ ,
\end{equation}
and the Maurer-Cartan current is
\begin{eqnarray}
  J & = & \frac{1}{y} (\partial x - 2 \partial \theta \bar \theta -
  \partial x \bar \theta \xi) P + \frac{1}{y^{1/2}} (\partial \theta
  - \partial x \xi +\partial \theta \bar \theta \xi) Q +
  \frac{1}{y^{1/2}} \partial \bar \theta \bar Q + y^{1/2} \partial \xi
  S + \nonumber \\
  && {} + \left( \frac{\partial y}{y} - \partial \bar \theta \xi -
  \partial \xi \bar \theta \right) D - \frac{i}{2} (\partial
  \bar \theta \xi + \partial \xi \bar \theta) R \ .
\end{eqnarray}

In this gauge the action is
\begin{eqnarray}
  S & = & \frac{R^2}{4 \pi \alpha'} \int d^2 z \Bigg[ \frac{(-1 + \bar
    \theta \xi) \partial x \bar \partial x + \partial y \bar \partial
    y}{2 y^2} - \frac{1}{y^2} \bar \theta ( \partial x \bar \partial
    \theta + \partial \theta \bar \partial x) - \nonumber \\
    && {} - \frac{1}{y} \xi (\partial x \bar \partial \bar \theta -
    \partial \bar \theta \bar \partial x) + \frac{1}{2 y} \xi
    (\partial y \bar \partial \bar \theta + \partial \bar
    \theta \bar \partial y) + \frac{1}{2 y} \bar \theta (\partial y
    \bar \partial \xi  + \partial \xi \bar \partial y) + \nonumber \\
    && {} + \frac{1}{2} \bar \theta \xi (\partial \bar \theta \bar
    \partial \xi - \partial \xi \bar \partial \bar \theta) +
    \frac{1}{y} (1 + \bar \theta \xi) (\partial \theta \bar \partial
    \bar \theta + \partial \bar \theta \bar \partial \theta) \Bigg]
  \ .
\end{eqnarray}
This action has no quadratic term in the Grassmann coordinate $\theta$
so the usual procedure introduced by Buscher
\cite{Buscher:1987sk,Buscher:1987qj} cannot be applied. This lends
credence to the explanation that in the different gauge above the
procedure failed because the quadratic term is $\kappa$-symmetry-exact.

\subsection{The $AdS_4$ target-space}

The non-critical superstring on $AdS_4$ with eight
supersymmetries can be constructed as the supercoset
$\mathrm{OSp}(2|4)/(\mathrm{SO}(3,1) \times \mathrm{SO}(2))$. Taking
$g \in \mathrm{OSp}(2|4)$ in (\ref{eq:GS-coset-action}) one obtains
the sigma-model
\begin{eqnarray}
  S & = & -\frac{R^2}{4 \pi \alpha'} \int d^2 z \Big[ \eta_{m n}
    (J_{P_m} + J_{K_m}) (\bar J_{P_n} + \bar J_{K_n}) + J_D \bar J_D +
    \nonumber \\
    && {} + 4 i \epsilon_{\alpha \beta} \left(J_{Q_\alpha} \bar J_{\hat
      Q_\beta} - J_{\hat Q_\alpha} \bar J_{Q_\beta} +J_{S_\alpha} \bar
    J_{\hat S_\beta} - J_{\hat S_\alpha} \bar J_{S_\beta} \right) \ ,
\end{eqnarray}
where the conventions for the algebra are given in Appendix
\ref{sec:osp24-algebra}.

The general group element can be parameterized as
\begin{equation}
  g = e^{x^m P_m + w^m K_m + \theta^\alpha Q_\alpha} e^B \ , \quad
  e^B = e^{\hat \theta^\alpha \hat Q_\alpha + \xi^\alpha S_\alpha +
    \hat \xi^\alpha \hat S_\alpha} y^D e^{\phi R + \omega^{m n} M_{m
      n}} \ .
\end{equation}
We assume that we can partially gauge-fix the $\kappa$-symmetry such that
$\hat \xi^\alpha = 0$ and we will also fix the $SO(3, 1) \times SO(2)$
gauge symmetry by setting $w^m = 0$, $\phi = 0$ and $\omega^{m n} = 0$
essentially picking the specific coset representative
\begin{equation}
  g = e^{x^m P_m + \theta^\alpha Q_\alpha} e^B \ , \quad
  e^B = e^{\hat \theta^\alpha \hat Q_\alpha + \xi^\alpha S_\alpha} y^D
  \ .
\end{equation}
One  can check that the needed components of the Maurer-Cartan 1-form
are
\begin{eqnarray}
  J_{P_m} & = & \left[ e^{-B} (dx^n P_n + d\theta^\alpha Q_\alpha) e^B
    \right]_{P_m} \ , \quad
  J_{K_m} = 0 \ , \nonumber \\
  J_{Q_\alpha} & = & \left[ e^{-B} (dx^m P_m + d\theta^\beta Q_\beta)
    e^B  \right]_{Q_\alpha}\ , \quad
  J_{S_\alpha} = \left[ e^{-B} de^B \right]_{S_\alpha} \ , \nonumber
  \\
  J_{\hat Q_\alpha} & = & \left[ e^{-B} de^B \right]_{\hat Q_\alpha}
  \ , \quad
  J_{\hat S_\alpha} = 0 \ , \quad
  J_D = \left[ e^{-B} de^B \right]_D \ .
\end{eqnarray}
The action then becomes
\begin{eqnarray}
  S & = & -\frac{R^2}{4 \pi \alpha'} \int d^2 z \Big[ \eta_{m n}
    \left[ e^{-B} (\partial x^l P_l + \partial \theta^\alpha Q_\alpha)
      e^B \right]_{P_m} \left[ e^{-B} ( \bar \partial x^p P_p + \bar
      \partial \theta^\beta Q_\beta) e^B \right]_{P_n} + \nonumber \\
    && {} + \left[ e^{-B} \partial e^B \right]_D \left[ e^{-B} \bar
      \partial e^B \right]_D + 4 i \epsilon_{\alpha \beta} \Big(
    \left[ e^{-B} (\partial x^m P_m + \partial \theta^\gamma Q_\gamma)
      e^B \right]_{Q_\alpha} \left[ e^{-B} \bar \partial e^B
      \right]_{\hat Q_\alpha} \nonumber \\
    && {} - \left[ e^{-B} \partial e^B
      \right]_{\hat Q_\alpha} \left[ e^{-B} ( \bar \partial x^m P_m + \bar
      \partial \theta^\gamma Q_\gamma) e^B \right]_{Q_\beta} \Big)
    \Big] \ .
\end{eqnarray}
Replacing the partial derivatives of $x^m$ and of $\theta^\alpha$ by
the gauge fields $A^m$, $\bar A^m$, $A^\alpha$ and $\bar A^\alpha$ and
then performing the field redefinition
\begin{eqnarray}
  A'^m & = & \left[ e^{-B} (A^n P_n + A^\alpha Q_\alpha) e^B
    \right]_{P_m} \ , \nonumber \\
  A'^\alpha & = & \left[ e^{-B} (A^m P_m + A^\beta Q_\beta) e^B
    \right]_{Q_\alpha}
\end{eqnarray}
with similar expressions for the right-moving sector and introducing the
Lagrange multiplier terms forcing the gauge fields to be flat, the new
equivalent action takes the form
\begin{eqnarray}
  S & = & -\frac{R^2}{4 \pi \alpha'} \int d^2 z \Big[ \eta_{m n} A'^m
    \bar A'^n + J_D \bar J_D + 4 i \epsilon_{\alpha \beta} \Big(
    A'^\alpha \left[ e^{-B} \bar \partial e^B \right]_{\hat Q_\beta} -
    \nonumber \\
    && {} - \left[ e^{-B} \partial e^B \right]_{\hat Q_\alpha} \bar
    A'^\beta \Big) + \tilde x_m (\bar \partial A^m - \partial \bar
    A^m) + \tilde \theta_\alpha (\bar \partial A^\alpha - \partial
    \bar A^\alpha) \Big] \ .
\end{eqnarray}
As can be immediately seen, just as for the case of the $AdS_2$ space
discussed in section \ref{sec:AdS2-sigma-model} for the (physically
trivial) $AdS_2$ the fermionic gauge fields $A'^\alpha$ and $\bar
A'^\alpha$ appear only linearly in the action and hence integration
over them will yield constraints rather than equations of motion.

\subsection{The $AdS_4 \times \mathbb{C}P^3$ target-space}
Using the same procedure as for the $AdS_n\times S^n$ models we construct the Green-Schwarz sigma model action (\ref{eq:GS-coset-action}) using the bilinear forms of (\ref{bilinear_forms_osp64}),
\be
S=\frac{R^2}{4 \pi \alpha'} \int d^2 z \Big[
-2\eta_{ab}(J_{P_a}+J_{K_a})(\bar J_{P_b}+\bar J_{K_b})
\ee
$$
-J_D\bar J_D
-2(J_{R_{kl}}\bar J_{R_{\dot p\dot q}}+\bar J_{R_{kl}}J_{R_{\dot p\dot q}})(\d^{k\dot p}\d^{l\dot q}-\d^{k\dot q}\d^{l\dot p})
$$
$$
-2i\d^{\dot k l}C_{\a\b}\Big(
 J_{Q^l_\a}\bar J_{Q^{\dot k}_\b}
-\bar J_{Q^l_\a}J_{Q^{\dot k}_\b}
-J_{S^l_\a}\bar J_{S^{\dot k}_\b}
+\bar J_{S^l_\a}J_{S^{\dot k}_\b}
\Big)
\Big] \ .
$$
(This coset model is a Green-Schwarz superstring with kappa-symmetry
partially fixed so that the broken supersymmetries are gauged away,
leaving only the fermionic coordinates corresponding to unbroken
supersymmetries. This partial gauge-fixing is not compatible with all
the string solutions so a general analysis would seem to require the
use of the complete unfixed sigma-model derived in
\cite{Gomis:2008jt}. However, as the broken supersymmetries are not
isometries of the background, they cannot participate in the T-duality
transformation so we expect the $\mathrm{OSp}(6|4)$ coset model to
suffice for this purpose.)

We proceed by taking the gauge-fixed coset representative
\be
g=e^{x^a P_a+\t^\a_l Q^l_\a}e^B, \quad
e^B\equiv e^{\t^\a_{\dot l} Q^{\dot l}_\a+\xi^\a_{ l} S^{ l}_\a}y^D e^{(\S y^{kl}R_{kl}+\S y^{\dot k\dot l}R_{\dot k\dot l})/y} \ ,
\ee
where we also fixed six out of the eight $\k$-symmetry degrees of
freedom by setting $\xi^\a_{\dot l}=0$. $P_a$ and $\t^\a_l Q^l_\a$
form an Abelian subalgebra.

Next, we would like to write the currents in terms of this parameterization, but in this case the commutation relations $[Q,R]\sim \bar Q$ and $\{Q,S\}\sim R$ prevent us from getting two kinds of currents as in (\ref{current_separation}), so
\be
J_{T}=[e^{-B}(dx^a P_a+d\t^\a_l Q^l_\a)e^{B}]_T+[e^{-B} d e^{B}]_T\equiv j_T+\mathfrak{j}_T \ .
\ee
We have
\be
j=j_{P_a}P_a+j_{Q^\a_l} Q^\a_l+j_{Q^\a_{\dot l}} Q^\a_{\dot l}+j_{R_{kl}}R_{kl}+j_{R_{\dot k\dot l}}R_{\dot k\dot l}+j_{\l_{\k\dot l}}\l_{\k\dot l} \ ,
\ee
and
\be
\mathfrak{j}
=\mathfrak{j}_{Q^\a_l} Q^\a_l+\mathfrak{j}_{Q^\a_{\dot l}} Q^\a_{\dot l}
+\mathfrak{j}_{S^\a_l} S^\a_l+\mathfrak{j}_{S^\a_{\dot l}} S^\a_{\dot l}+\mathfrak{j}_{R_{kl}}R_{kl}+\mathfrak{j}_{R_{\dot k\dot l}}R_{\dot k\dot l}
+\mathfrak{j}_{\l_{\k\dot l}}\l_{\k\dot l}+\mathfrak{j}_{D}D+\mathfrak{j}_{M_{mn}}M_{mn} \ .
\ee
The action in terms of these currents reads
\be
S=\frac{R^2}{4 \pi \alpha'} \int d^2 z \Big[
-2\eta^{ab}j_{P_a}\bar j_{P_b}
-\mathfrak{j}_D\bar \mathfrak{j}_D
\ee
$$
-2((j+\mathfrak{j})_{R_{kl}}(\bar j+\bar \mathfrak{j})_{R_{\dot p\dot q}}
+(\bar j+\bar \mathfrak{j})_{R_{kl}}(j+\mathfrak{j})_{R_{\dot p\dot q}})(\d^{k\dot p}\d^{l\dot q}-\d^{k\dot q}\d^{l\dot p})
$$
$$
-2i\d^{\dot k l}C_{\a\b}\Big(
 (j+\mathfrak{j})_{Q^l_\a}(\bar j+\bar \mathfrak{j})_{Q^{\dot k}_\b}
-(\bar j+\bar \mathfrak{j})_{Q^l_\a}(j+\mathfrak{j})_{Q^{\dot k}_\b}
-\mathfrak{j}_{S^l_\a}\mathfrak{j}_{S^{\dot k}_\b}
+\mathfrak{j}_{S^l_\a}\mathfrak{j}_{S^{\dot k}_\b}
\Big)
\Big] \ .
$$
We add gauge fields instead of the $x_a$ and $\t^\a_l$ derivatives and a suitable Lagrange multiplier term. We define
\be\label{eq:A-def-osp64}
A'=e^{-B}(A^a P_a+A^\a_l Q^l_\a)e^{B} \ ,
\ee
(where $A^a\equiv A_{P_a}$ and $A^\a_l\equiv A_{Q^l_\a}$) and the action becomes
\be
S=\frac{R^2}{4 \pi \alpha'} \int d^2 z \Big[
-2\eta_{ab}A'_{P_a}\bar A'_{P_b}
-\mathfrak{j}_D\bar \mathfrak{j}_D
\ee
$$
-2((A'+\mathfrak{j})_{R_{kl}}(\bar A'+\bar \mathfrak{j})_{R_{\dot p\dot q}}
+(\bar A'+\bar \mathfrak{j})_{R_{kl}}(A'+\mathfrak{j})_{R_{\dot p\dot q}})(\d^{k\dot p}\d^{l\dot q}-\d^{k\dot q}\d^{l\dot p})
$$
$$
-2i\d^{\dot k l}C_{\a\b}\Big(
 (A'+\mathfrak{j})_{Q^l_\a}(\bar A'+\bar \mathfrak{j})_{Q^{\dot k}_\b}
-(\bar A'+\bar \mathfrak{j})_{Q^l_\a}(A'+\mathfrak{j})_{Q^{\dot k}_\b}
-\mathfrak{j}_{S^l_\a}\mathfrak{j}_{S^{\dot k}_\b}
+\mathfrak{j}_{S^l_\a}\mathfrak{j}_{S^{\dot k}_\b}
\Big)
$$
$$
+\tilde x^a(\p \bar A_a-\bar\p A_a)
+\tilde \t^\a_l(\p \bar A^l_\a-\bar\p A^l_\a)
\Big] \ .
$$
\subsubsection{Current expansion in fermions}
To zeroth order in the fermions ($\t^\a_l,\t^\a_{\dot l},\xi^\a_l$) we get a bosonic sigma model with no fermions, which we can T-dualize,
\be
S_0=\frac{R^2}{4 \pi \alpha'} \int d^2 z \Big[
-2y^2\eta_{ab}A_{P_a}\bar A_{P_b}
-j_D\bar j_D
\ee
$$
-2\bigg\{\mathfrak{j}_{R_{kl}}\bar \mathfrak{j}_{R_{\dot p\dot q}}
+\bar \mathfrak{j}_{R_{kl}}\mathfrak{j}_{R_{\dot p\dot q}}\bigg\}(\d^{k\dot p}\d^{l\dot q}-\d^{k\dot q}\d^{l\dot p})
+\tilde x^a(\p \bar A_a-\bar\p A_a)
\Big] \ ,
$$
where $\mathfrak{j}=e^{-B}de^B$, $e^B=y^D e^{(\S y^{kl}R_{kl}+\S y^{\dot k\dot l}R_{\dot k\dot l})/y}$.

Next, leaving only $\t^\a_l$ terms we find
\be
e^{-B}(A^a P_a + A^\a_\l Q^l_\a)e^{B}=yA^a P_a+y^{1/2}A^\a_k\left(f^k_l(\frac{y^{mn}}{y},\frac{y^{\dot m\dot n}}{y}) Q^l_\a +g^k_{\dot l}(\frac{y^{mn}}{y},\frac{y^{\dot m\dot n}}{y}) Q^{\dot l}_\a\right) \ ,
\ee
thus the $A'$'s in terms of the $A$'s are
\be
A'^a=yA^a,\quad
A'^\a_{l}=y^{1/2}A^\a_k f^k_l(\frac{y^{mn}}{y},\frac{y^{\dot m\dot n}}{y}),\quad
A'^\a_{\dot l}=y^{1/2}A^\a_k g^k_{\dot l}(\frac{y^{mn}}{y},\frac{y^{\dot m\dot n}}{y}) \ .
\ee
Plugging these in the action  we have
\be\label{eq:first_order_action}
S_1=\frac{R^2}{4 \pi \alpha'} \int d^2 z \Big[
-2y^2\eta_{ab}A_{P_a}\bar A_{P_b}
-\mathfrak{j}_D\bar \mathfrak{j}_D
-2\bigg\{\mathfrak{j}_{R_{kl}}\bar \mathfrak{j}_{R_{\dot p\dot q}}
+\bar \mathfrak{j}_{R_{kl}}\mathfrak{j}_{R_{\dot p\dot q}}\bigg\}(\d^{k\dot p}\d^{l\dot q}-\d^{k\dot q}\d^{l\dot p})
\ee
$$
-2iy\d^{\dot k l}C_{\a\b}\Big(
 A^\a_n f^n_l\bar A^\b_m g^m_{\dot k}
-\bar A^\a_n f^n_l A^\b_m g^m_{\dot k}
\Big)
+\tilde x^a(\p \bar A_a-\bar\p A_a)
+\tilde \t^\a_l(\p \bar A^l_\a-\bar\p A^l_\a)
\Big]
$$
$$
=\frac{R^2}{4 \pi \alpha'} \int d^2 z \Big[
-2y^2\eta_{ab}A_{P_a}\bar A_{P_b}
-\mathfrak{j}_D\bar \mathfrak{j}_D
-2\bigg\{\mathfrak{j}_{R_{kl}}\bar \mathfrak{j}_{R_{\dot p\dot q}}
+\bar \mathfrak{j}_{R_{kl}}\mathfrak{j}_{R_{\dot p\dot q}}\bigg\}(\d^{k\dot p}\d^{l\dot q}-\d^{k\dot q}\d^{l\dot p})
$$
$$
-2iy\d^{\dot k l}C_{\a\b} A^\a_n\bar A^\b_m \Big(
 f^n_lg^m_{\dot k}
- f^m_lg^n_{\dot k}
\Big)
+\tilde x^a(\p \bar A_a-\bar\p A_a)
+\tilde \t^\a_l(\p \bar A^l_\a-\bar\p A^l_\a)
\Big] \ .
$$
The bosonic T-duality works as before, but now we also have a quadratic
term for the fermions. The equation of motion for the fermions is
\be\label{eq:first_order_constraint}
A^\a_m \d^{l\dot k}\left(
f^n_l g^m_{\dot k}-f^m_l g^n_{\dot k}
\right)
=-\frac{i}{2y}C^{\a\b}\p\t^n_\b
\ .
\ee
The matrix $M^{mn}\equiv \d^{l\dot k}\left(
f^n_l g^m_{\dot k}-f^m_l g^n_{\dot k}
\right)=-4iy^{\dot n \dot m}/y+O(y^{\dot n \dot m}y^{kl}/y^2)$,
is an antisymmetric three-dimensional matrix and hence has a vanishing determinant, so we cannot solve for $A^\a_l$.

Next, to first order in $\xi^\a_l$ we have nontrivial $\mathfrak{j}_{S^l_\a}$ and $\mathfrak{j}_{S^{\dot l}_\a}$,
but these terms do not mix with the $A$'s so we will ignore them. The $A$'s change as follows,
\be
A^\b_k\rightarrow A^\b_k-iA^a(\g_a)_\a{}^\b\xi^\a_k
\ee
$$
A^a\rightarrow A^a
$$
$$
A_{kl}\rightarrow A^\b_k C_{\b\a}\xi^\a_l,\quad
A'^{rs}\rightarrow h^{rs}_{kl}(y^{ij}/y)A^\b_k C_{\b\a}\xi^\a_l \ .
$$
Thus the change in (\ref{eq:first_order_constraint}) goes like
\be\label{eq:second_order_constraint}
(A^\a_m-iA^a(\g_a)_\b{}^\a\xi^\b_m)\d^{l\dot k}\left(
f^n_l g^m_{\dot k}-f^m_l g^n_{\dot k}
\right)
=-\frac{i}{2y}C^{\a\b}\p\t^n_\b+k^{\a n}(\xi,\mathfrak{j}_{R_{kl}},y,...) \ ,
\ee
and again, we have the same singular matrix $M^{mn}$ multiplying the gauge fields $A$.
The second equation, which we can think of as the first order correction to the bosonic T-duality equation, is
\be
2y^2 A^a
+2\d^{\dot k l} \xi^\g_l (\g^a C)_{\g\b}(\sqrt{y} A^\b_m g^m_{\dot k}+\mathfrak{j}^\b_{\dot k})=-\p \tilde x^a \ .
\ee
We can plug the solution for $A^m$ in
(\ref{eq:second_order_constraint}), but we will just get an expression
of the form
\be
A^\b_l(\d^\a_\b \d ^l_m +\phi^{\a}_{\b}F^l_m)M^{mn}=-\frac{i}{2y}C^{\a\b}\p\t^n_\b+k^{\a n}(\xi,\mathfrak{j}_{R_{kl}},y,...)+...
\ee
where $\phi$, $F$, $k$ and $...$ are functions of the coordinates but not of $A$, so again we cannot solve for $A^\b_l$.

Going next to first order in $\t^\a_{\dot l}$ (but leaving out terms of order $\t^\a_{\dot l}\xi^\b_{k}$) we have nontrivial $\mathfrak{j}_{Q^l_\a}$ and $\mathfrak{j}_{Q^{\dot l}_\a}$, and also
\be
A^a\rightarrow A^a+A_k^\b\d^{\dot l k}(\g^a C)_{\b\a}\t^\a_{\dot l}
\ee
with respect to (\ref{eq:first_order_action}), so the action we get is
\be\label{eq:second_order_action}
S_2=\frac{R^2}{4 \pi \alpha'} \int d^2 z \Big[
-2y^2\eta_{ab}(A^a+A_k^\b\d^{\dot l k}(\g^a C)_{\b\a}\t^\a_{\dot l})(\bar A^b+\bar A_m^\g\d^{\dot n m}(\g^b C)_{\g\d}\t^\d_{\dot n})
\ee
$$
-\mathfrak{j}_D\bar \mathfrak{j}_D
-2\bigg\{(h^{kl}_{rs}A^\b_r C_{\b\a}\xi^\a_s+\mathfrak{j}_{kl})\bar \mathfrak{j}_{\dot p\dot q}
+(h^{kl}_{mn}\bar A^\b_m C_{\b\a}\xi^\a_n+\bar \mathfrak{j}_{kl})\mathfrak{j}_{\dot p\dot q}\bigg\}(\d^{k\dot p}\d^{l\dot q}-\d^{k\dot q}\d^{l\dot p})
$$
$$
-2i\d^{\dot k l}C_{\a\b}\Big(
y(f^m_l(A^\a_m-iA^a(\g_a)_\d{}^\a\xi^\d_m)+\mathfrak{j}^\a_l)
(g^n_{\dot k}(\bar A^\b_n-i\bar A^b(\g_b)_\g{}^\b\xi^\g_n)+\bar \mathfrak{j}^\b_{\dot k})
$$
$$
-y(f^m_l(\bar A^\a_m-i\bar A^a(\g_a)_\d{}^\a\xi^\d_m)+\bar \mathfrak{j}^\a_l)
(g^n_{\dot k}( A^\b_n-i A^b(\g_b)_\g{}^\b\xi^\g_n)+ \mathfrak{j}^\b_{\dot k})
-\mathfrak{j}_{S^l_\a}\mathfrak{j}_{S^{\dot k}_\b}
+\mathfrak{j}_{S^l_\a}\mathfrak{j}_{S^{\dot k}_\b}
\Big)
$$
$$
+\tilde x^a(\p \bar A_a-\bar\p A_a)
+\tilde \t^\a_l(\p \bar A^l_\a-\bar\p A^l_\a)
\Big] \ .
$$
The equations of motion (to first order in $\xi$ or $\t$) are,
\be\label{eq:bosoneq1}
-2yA_a+A^\b_k\d^{\dot l k}(\g_a C)_{\b\a}\t^\a_{\dot l}
\ee
$$
-2\d^{\dot k l}y C_{\a\b}\Big[
g^n_{\dot k}(f^m_l A^\a_m +\mathfrak{j}^\a_l)(\g_a)_\g{}^{\b}\xi^\g_n
-f^m_l(\g_a)_\d{}^\a\xi^\d_m(g^n_{\dot k}A^\b_n+\mathfrak{j}^\b_{\dot k})
\Big]=\p\tilde x_a
$$
\be\label{eq:fermioneq1}
2i\d^{\dot k l}C_{\a\b}y\Big\{
(f^m_l(A^\a_m-iA^a(\g_a)_\d{}^\a\xi^\d_m)+\mathfrak{j}^\a_l)g^n_{\dot k}
-f^n_l(g^m_{\dot k}(A^\a_m-iA^a(\g_a)_\g{}^\a\xi^\g_m)+\mathfrak{j}^\a_{\dot k})
\Big\}
\ee
$$
-2yA^a\d^{n\dot m}(\g_a C)_{\b\d}\t^\d_{\dot m}+F^n_\b=0 \ .
$$
Plugging (\ref{eq:bosoneq1}) in (\ref{eq:fermioneq1}) we find
\be
2i\d^{\dot k l}C_{\a\b}y\Big\{
(f^m_l(A^\a_m+i\frac{\p\tilde x^a}{2y}(\g_a)_\d{}^\a\xi^\d_m)+\mathfrak{j}^\a_l)g^n_{\dot k}
-f^n_l(g^m_{\dot k}(A^\a_m+i\frac{\p\tilde x^a}{2y}(\g_a)_\g{}^\a\xi^\g_m)+\mathfrak{j}^\a_{\dot k})
\Big\}
\ee
$$
+\p\tilde x^a\d^{n\dot m}(\g_a C)_{\b\d}\t^\d_{\dot m}+F^n_\b=0 \ .
$$
Thus again the $A^\a_l$ will multiply the same singular matrix.

More generally, the bosonic singular matrix $M$ will get corrections with even powers of the fermionic variables,
\be
M_{mn}=M^{s}_{mn}+M^{2}_{mn}+M^{4}_{mn}+... \ ,
\ee
where $M^{2n}_{mn}$ is a matrix function with a power of $2n$ of the
fermionic variables $\t^\a_{\dot k}$ and/or $\xi^\a_k$, and
$M^{s}_{mn}$ is the purely bosonic singular matrix. In order to solve
the equations of motion we will need the inverse of $M_{mn}$, but such an inverse does not exists, because all the matrices involving fermions should cancel each other order by order and we should get,
\be
\d_{mn}=M_{mk}M^{-1}_{kn}=M^s_{mk}(M^{-1})^s_{kn}+...=M^s_{mk}(M^{-1})^s_{kn} \ ,
\ee
but $M^{s}_{mn}$ is singular.

\subsection{The $AdS_2$ target-space with eight supersymmetries}
Similarly to section \ref{sec:AdS_2xS^2_target_space} we construct a
Green-Schwarz sigma-model action on $AdS_2$, but using a different
supercoset, namely $\mathrm{PSU}(1,1|2)/(\mathrm{U}(1)\times
\mathrm{SU}(2))$, so this time we will have eight supersymmetries rather then four. Using the $\mathbb{Z}_4$ structure of this super-algebra given in (\ref{eq:Z4structure_psu112_AdS2}) we construct the Green-Schwarz action
\be
  S = \frac{R^2}{4 \pi \a'}
   \int d^2 z  \left( -J_P \bar J_K - J_K \bar J_P
  +\frac{i}{2}\e_{\a\b}( J_{Q_\a} \bar J_{S_\b} -J_{\hat Q_\a} \bar J_{\hat S_\b}-J_{S_\a} \bar J_{Q_\b} + J_{\hat S_\a} \bar J_{\hat Q_\b}\right) \ .
\ee
We parameterize the coset representatives such that
\be
g=e^{x P+ \t^\a Q_\a}e^B,\qquad
e^B=e^{\hat\t^\a \hat Q_\a+\xi^\a S_\a+\hat\xi^\a \hat S_\a}e^{y K} \ .
\ee
If we define the following currents
\be
J=g^{-1}dg=j+\mathfrak{j},\quad
j=e^{-B}(dxP+d\t^\a Q_\a)e^B, \qquad
\mathfrak{j}=e^{-B} d e^B, \qquad
\ee
Expanding $j$ to zeroth order in the fermions in $B$ we get,
\be
j_{(0)}=dx P+s\t^\a Q_\a+2y dx D+id\t^\a \hat S_\a-y^2 dx K \ .
\ee
Thus by using the Buscher procedure of introducing gauge fields as in
the examples above, to this order the action will not have quadratic terms in the fermions, so all the quadratic $d\t$ (or more precisely $A^\a$) terms in the action,
coming from higher orders in the fermions in $B$, will multiply some
fermions, and we will not be able to solve the equations of motion for the fermions.
The reason again is that we will have to multiply the equations of motion by a singular matrix.
%

\section{A general analysis}
In this section we present a general algebraic argument to when a supercoset is expected to have a
fermionic T-duality symmetry, and when it will fail to have one.

Let us assume a superconformal algebra $\mathcal{G}$ with a
$\mathbb{Z}_4$ automorphism structure with the zero grading subalgebra
$\mathcal{H}$, with the usual conformal bosonic generators $P_\m$,
$K_\m$, $D$, $J_{\m\n}$, supercharges $Q$, $\bar Q$, superconformal
generators $S$, $\bar S$ and R-symmetry generators $R_a$. We can find
an Abelian subalgebra $\mathcal{A}$ composed of a subset of the $P$'s
and $Q$'s if we can find an  anti-commuting combination of supercharges.
We denote by $A,B,..$ the indices of $\mathcal{A}$. Uppercase letters
denote both bosonic and fermionic indices, lowercase letters bosonic
ones and Greek letters fermionic ones, $A=\{a,\a\}$. We gauge fix the
$\mathcal{H}$ symmetry and parameterize the coset element as follows,
\be
g=e^{x_A T^A}e^B, \quad x_A T^A\in \mathcal{A},\quad B\in \mathcal{G}\ominus\mathcal{A} \ .
\ee
More specifically we parameterize
\be
e^B=\exp(\bar\t \bar Q+\xi S+\bar\xi \bar S)y^D \exp(y^\m R_\m /y) \ ,
\ee
where contraction of the fermions is understood ($R_\m$ includes only the generators in the supercoset, $R_\m\in \mathcal{G}/\mathcal{H}$).
The left-invariant one-form current will split into two pieces,
\be\label{eq:leftinvariant-oneform}
J=e^{-B} dx_a T^a e^B + e^{-B} d e^B\equiv j+\mathfrak{j} \ ,
\ee
where $\mathfrak{j}$ is independent of the coordinates associated with the Abelian subalgebra $x_a$. Generally, $j$ and $\mathfrak{j}$ can take any value in $\mathcal{G}$. We give special indices to these generators as follows,
\be
j=j_I T^I,\quad \mathfrak{j}=\mathfrak{j}_W T^W
\ee
(in general $a$ is both in $I$ and $W$ and $T^I \cap T^W\neq\varnothing$), so we can write the general Green-Schwarz sigma model action as,
\be
S=\int d^2 z\Big[
 j^{(2)}_I \bar j^{(2)}_J \eta^{IJ}
+ \mathfrak{j}^{(2)}_W \bar j^{(2)}_J \eta^{W J}
+ j^{(2)}_I \bar \mathfrak{j}^{(2)}_X \eta^{IX}
+ \mathfrak{j}^{(2)}_W \bar \mathfrak{j}^{(2)}_X \eta^{WX}
\ee
$$
+\half\Big( j^{(1)}_I \bar j^{(3)}_J \eta^{IJ}
+ \mathfrak{j}^{(1)}_W \bar j^{(3)}_J \eta^{W J}
+ j^{(1)}_I \bar \mathfrak{j}^{(3)}_X \eta^{IX}
+ \mathfrak{j}^{(1)}_W \bar \mathfrak{j}^{(3)}_X \eta^{WX}
\Big)
$$
$$
-\half\Big( \bar j^{(1)}_I j^{(3)}_J \eta^{IJ}
+  \bar \mathfrak{j}^{(1)}_W j^{(3)}_J \eta^{W J}
+  \bar j^{(1)}_I \mathfrak{j}^{(3)}_X \eta^{IX}
+  \bar \mathfrak{j}^{(1)}_W \mathfrak{j}^{(3)}_X \eta^{WX}
\Big)
\Big]
$$
$$
=\int d^2 z\Big[
 A'^{(2)}_I \bar A'^{(2)}_J \eta^{IJ}
+ \mathfrak{j}^{(2)}_W \bar A'^{(2)}_J \eta^{W J}
+ A'^{(2)}_I \bar \mathfrak{j}^{(2)}_X \eta^{IX}
+ \mathfrak{j}^{(2)}_W \bar \mathfrak{j}^{(2)}_X \eta^{WX}
$$
$$
+\half\Big( A'^{(1)}_I \bar A'^{(3)}_J \eta^{IJ}
+ \mathfrak{j}^{(1)}_W \bar A'^{(3)}_J \eta^{W J}
+ A'^{(1)}_I \bar \mathfrak{j}^{(3)}_X \eta^{IX}
+ \mathfrak{j}^{(1)}_W \bar \mathfrak{j}^{(3)}_X \eta^{WX}
\Big)
$$
$$
-\half\Big( \bar A'^{(1)}_I A'^{(3)}_J \eta^{IJ}
+  \bar \mathfrak{j}^{(1)}_W A'^{(3)}_J \eta^{W J}
+  \bar A'^{(1)}_I \mathfrak{j}^{(3)}_X \eta^{IX}
+  \bar \mathfrak{j}^{(1)}_W \mathfrak{j}^{(3)}_X \eta^{WX}
\Big)
$$
$$
+\tilde x^A (\p \bar A_A-\bar \p A_A)
\Big] \ ,
$$
where $A'$ equals $j$ when replacing the coordinate derivatives $dx_A$ with the gauge field $A_A$, and $\tilde x_A$ is the Lagrange multiplier.
By the structure of $j$, (\ref{eq:leftinvariant-oneform}), we see that $A'$ depends linearly on $A$, so quadratic terms will rise only from $A'\bar A'$ interactions in the action.
We also know that
\be
A'^{(1)}=Q^I A'^{(1)}_{Q^I}+\bar Q^I A'^{(1)}_{\bar Q^I}+S^I A'^{(1)}_{S^I}+\bar S^I A'^{(1)}_
{\bar S^I} \ ,
\ee
and similarly for $A'^{(3)}$.
Generally, we encountered two cases where $J_{Q^I}$ is coupled to $\bar J_{Q^I}$ or to $\bar J_{\bar Q^I}$, namely we have terms in the action of the form \be
c\eta^{IJ}(J_{Q^I}\bar J_{Q^J}-J_{\bar Q^I}\bar J_{\bar Q^J}), \quad \mathrm{or}\quad
c\eta^{IJ}(J_{Q^I}\bar J_{\bar Q^J}-\bar J_{Q^I}J_{\bar Q^J}) \ ,
\ee
where $c$ is some constant. Let us call these two cases, case I and
case II respectively.  Case I appeared usually when considering
$\mathrm{PSU}$ based model and case II when considering  ortho-symplectic based model.
To zeroth order in the fermions of $B$ we have
\be\label{eq:Aprimezerothtrans}
A'_i=f(y)A_i,\quad
A'_{Q^\iota}=g(y)h_\iota^\k(y^\m/y) A_{Q^\k},\quad
A'_{\bar Q^\iota}=g(y)\tilde h_{\iota}^{\k}(y^\m/y) A_{\bar Q^\k},\quad
\ee
where the $\iota,\k$ indices are the R-symmetry indices of $Q,\bar Q$, so $Q$ might have more indices of transformation under the bosonic conformal group, which are the same on both sides of the equations (\ref{eq:Aprimezerothtrans}).
To this order, the equations of motion for the bosons are
\be
A^{(2)}_I\eta^{IJ}f^2=F^J(\mathfrak{j},\tilde x^A) \ ,
\ee
while for the fermions we have for the two cases
\be
c\eta^{\iota \varsigma}g^2 h^\k_\varsigma h^\l_\iota A^{(1)}_\l=F^\k(\mathfrak{j},\tilde x^A),\quad \mathrm{and}\quad
c\eta^{\iota \varsigma}g^2(\tilde h^\k_\varsigma h^\l_\iota -\tilde h^\l_\varsigma h^\k_\iota)A^{(1)}_\l=G^\k(\mathfrak{j},\tilde x^A) \ ,
\ee
respectively. The matrix $m^{\k\l}\equiv\eta^{\iota \varsigma}\tilde
h^{[\k}_\varsigma h^{\l]}_\iota$ in the second equation is singular
when the dimension of the representation of $Q$ under R-symmetry is
odd. That is the case for the $AdS_4\times CP^3$ action.
The matrix $m$ will get corrections at higher orders in the fermions of $B$, but these will contribute additively with even number of fermions to keep $m$ bosonic. Let us call the full matrix $M$ so,
\be
M=m+\mathcal{O}(\chi^2) \ ,
\ee
where $\chi=\{\bar \t,\xi,\bar \xi\}$. The inverse matrix should have
the form $M^{-1}=m^{-1}+\mathcal{O}(\chi^2)$, since all the fermions
should cancel, so when $m$ is singular so is $M$.

In case II when the R-symmetry does not mix $Q$ with $\bar Q$ (as in
the $AdS_4$ case), we don't get a quadratic term to this
order. Actually in this case we will get quadratic terms in $A_Q$ when
going to higher orders, so we will get equations of motion for $A_Q$,
but these will always multiply terms of order $O(\chi^2)$ or higher in
the fermions, so again we will have a singular matrix multiplying
$A_Q$, and we will not be able to solve for them. The argument above
does not rely on fixing any fermionic degrees of freedom (specifically
we don't use $\k$-symmetry).

We saw that there can also be a case III where the WZ term gives an interaction of the form
\be
J_Q\bar J_S+... \ ,
\ee
and this is the case of $AdS_2$ with eight supersymmetries realized by
the supercoset $\mathrm{PSU}(1,1|2)/(\mathrm{U}(1)\times \mathrm{SU}(2))$.
In this case,
like for $AdS_4$, $A_S$ does not get corrections to zeroth order in
the fermions, so the quadratic term $\p\t \bar\p\t$ again will be
multiplied by a singular fermionic matrix. This case has a different
structure since the R-symmetry does not play any role (its generators are in $\mathcal{H}$), instead we have $K$ (special conformal transformations) that mixes $Q$ with $\bar S$ and $\bar Q$ with $S$.

\acknowledgments 
We would like to thank N.~Berkovits, P.~A.~Grassi, M.~Magro,
L.~Mazzucato, V.~Scho\-merus, D.~Sorokin and S.~Theisen for valuable
discussions. I.A.\ would like to thank the KITP, UCSB for hospitality
during the Fundamental Aspects of String Theory workshop. This
research was supported by the German-Israeli Project cooperation (DIP
H.52), the German-Israeli Fund (GIF) and supported in part by the
National Science Foundation under Grant No.~NSF~PHY05-51164.

\appendix

\section{The $\mathrm{osp}(2|2)$ algebra}  \label{sec:osp22-algebra}

The $\mathrm{osp}(2|2)$ algebra is
\begin{eqnarray}
  [P, K] & = & -2 D \ , \quad
  [D, P] = P \ , \quad
  [D, K] = -K \ , \quad
  [R, D] = 0 \ , \quad
  [R, P] = 0 \ , \nonumber \\
  {}[R, K] & = & 0 \ , \nonumber \\
  {}[P, Q] & = & 0 \ , \quad
  [P, \bar Q] = 0 \ , \quad
  [K, Q] = S \ , \quad
  [K, \bar Q] = \bar S \ , \quad
  [D, Q] = \frac{1}{2} Q \ , \nonumber \\
  {}[D, \bar Q] & = & \frac{1}{2} \bar Q \ , \quad
  [R, Q] = -i Q \ , \quad
  [R, \bar Q] = i \bar Q \ , \nonumber \\
  {}[P, S] & = & -Q \ , \quad
  [P, \bar S] = - \bar Q \ , \quad
  [K, S]= 0 \ , \quad
  [K, \bar S] = 0 \ , \quad
  [D, S] = -\frac{1}{2} S \ , \nonumber \\
  {}[D, \bar S] & = & -\frac{1}{2} \bar S \ , \quad
  [R, S] = -i S \ , \quad
  [R, \bar S] = i \bar S \ , \nonumber \\
  \{Q, Q\} & = & 0 \ , \quad
  \{ \bar Q, \bar Q \} = 0 \ , \quad
  \{ Q, \bar Q \} = 2 P \ , \quad
  \{ S, S \} = 0 \ , \quad
  \{ \bar S, \bar S \} = 0 \ , \nonumber \\
  \{ S, \bar S \} & = &2 K \ , \quad
  \{ Q, S \} = 0 \ , \quad
  \{ \bar Q, \bar S \} = 0 \ , \nonumber \\
  \{ Q, \bar S \} & = & 2 D - i R \ , \quad
  \{\bar Q, S \} = 2 D + i R \ .
\end{eqnarray}

This algebra admits a $\mathbb{Z}_4$ automorphism. The automorphism
relevant to our needs is the one implemented by $\Omega X \Omega^{-1}$
for $X \in \mathrm{osp}(2|2)$, where
\begin{equation}
  \Omega = \left(
  \begin{array}{cc}
    -i \sigma_1 & 0 \\
    0 & \mathbf{1}
  \end{array}
  \right) \ .
\end{equation}
Using this automorphism the algebra can be decomposed into
$\mathbb{Z}_4$-invariant subspaces $\mathcal{H}_k$ ($k = 0 \dots, 3$)
such that
\begin{displaymath}
  \mathcal{H}_k = \{ X \in \mathrm{osp}(2|2) | \Omega X \Omega^{-1} =
  i^k X \} \ .
\end{displaymath}
These subspaces are
\begin{eqnarray}
  \mathcal{H}_0 & = & \{ P - K, \ R \} \ , \\
  \mathcal{H}_1 & = & \{ Q + S, \ \bar Q + \bar S \} \ , \\
  \mathcal{H}_2 & = & \{ P + K, \ D \} \ , \\
  \mathcal{H}_3 & = & \{ Q - S , \ \bar Q - \bar S \} \ .
\end{eqnarray}
The Cartan-Killing bilinear form is defined by $g_{X Y} = \Str (X
Y)$. Its non-trivial elements are
\begin{eqnarray}
  \Str (P K) & = & \Str (K P) = -1 \ , \quad \Str (D D) = \frac{1}{2}
  \ , \quad
  \Str (R R) = 2 \ , \nonumber \\
  \Str (Q \bar S) & = & \Str (\bar Q S) = 2 \ .
\end{eqnarray}

\section{The $\mathrm{osp}(2|4)$ algebra} \label{sec:osp24-algebra}

The definition of the $\mathrm{OSp}(2|4)$ group is that of
\cite{Frappat:1996pb}. The non-trivial brackets of the
$\mathrm{osp}(2|4)$ algebra are
\begin{eqnarray}
  [M_{m n}, M_{p q}] & = & \eta_{m p} M_{n q} + \eta_{n q} M_{m p} -
  \eta_{m q} M_{n p} - \eta_{n p} M_{m q} \ , \nonumber \\
  {}[M_{m n}, P_p] & = & \eta_{m p} P_n - \eta_{n p} P_m \ , \quad
  [M_{m n}, K_p] = \eta_{m p} K_n - \eta_{n p} K_m \ , \nonumber \\
  {}[D, P_m] & =& P_m \ , \quad
  [D, K_m] = -K_m \ , \quad
  [P_m, K_n] = 2 \eta_{m n} D - 2 M_{m n} \ , \nonumber \\
  {}[D, Q_\alpha] & = & \frac{1}{2} Q_\alpha \ , \quad
  [D, \hat Q_\alpha] = \frac{1}{2} \hat Q_\alpha \ , \quad
  [D, S_\alpha] = -\frac{1}{2} S_\alpha \ , \quad
  [D, \hat S_\alpha] = -\frac{1}{2} \hat S_\alpha \ , \nonumber \\
  {}[M_{m n}, Q_\alpha] & = & \frac{1}{2} {(C \gamma_{m n}
    C^{-1})_\alpha}^\beta Q_\beta \ , \quad
  [M_{m n}, \hat Q_\alpha] = \frac{1}{2} {(C \gamma_{m n}
    C^{-1})_\alpha}^\beta \hat Q_\beta \ , \nonumber \\
  {}[M_{m n}, S_\alpha] & = & \frac{1}{2} {(C \gamma_{m n}
    C^{-1})_\alpha}^\beta S_\beta \ , \quad
  [M_{m n}, \hat S_\alpha] = \frac{1}{2} {(C \gamma_{m n}
    C^{-1})_\alpha}^\beta \hat S_\beta \ , \nonumber \\
  {}[P_m, S_\alpha] & = & i {(C \gamma_m C^{-1})_\alpha}^\beta Q_\beta
  \ , \quad
  [P_m, \hat S_\alpha] = i {(C \gamma_m C^{-1})_\alpha}^\beta \hat Q_\beta
  \ , \nonumber \\
  {}[K_m, Q_\alpha] & = & -i {(C \gamma_m C^{-1})_\alpha}^\beta
  S_\beta \ , \quad
  [K_m,\hat Q_\alpha] = -i {(C \gamma_m C^{-1})_\alpha}^\beta
  \hat S_\beta \ , \nonumber \\
  {}[R, Q_\alpha] & = & i Q_\alpha \ , \quad
  [R, \hat Q_\alpha] = -i \hat Q_\alpha \ , \quad
  [R, S_\alpha] = i S_\alpha \ , \quad
  [R, \hat S_\alpha] = -i \hat S_\alpha \nonumber \\
  \{Q_\alpha, \hat Q_\beta\} & = & 4 (C \gamma^m)_{\alpha \beta} P_m
  \ , \quad
  \{S_\alpha, \hat S_\beta\} = -4 (C \gamma^m)_{\alpha \beta} K_m \ ,
  \nonumber \\
  \{Q_\alpha, \hat S_\beta\} & = & -2 i (C \gamma^{m n})_{\alpha
    \beta} M_{m n} + 4 i \epsilon_{\alpha \beta} D - 4
  \epsilon_{\alpha \beta} R \ , \nonumber \\
  \{\hat Q_\alpha, S_\beta\} & = & -2 i(C \gamma^{m n})_{\alpha \beta}
  M_{m n} + 4 i \epsilon_{\alpha \beta} D + 4 \epsilon_{\alpha \beta} R
  \ ,
\end{eqnarray}
where $\epsilon_{12} = -\epsilon_{21} = 1$, $\epsilon^{21} =
-\epsilon^{12} = 1$, $C_{\alpha \beta} = \epsilon_{\alpha \beta}$ and
the Dirac matrices have the index structure ${\gamma^{m
    \alpha}}_\beta$ and indices are lowered and raised using $C$. The
antisymmetrized Dirac-matrices are $\gamma^{m n} = \frac{1}{2}
[\gamma^m, \gamma^n]$. The metric is $\eta = \mathrm{diag}(-1, 1, 1)$,
$m,n,p,q = 0, \dots, 2$ are space-time indices, $\alpha, \beta = 1, 2$
are spinor indices and $I, J = 1, 2$ are $\mathrm{SO}(2)$ R-symmetry
indices.

The non-trivial elements of the Cartan-Killing bilinear form are
\begin{eqnarray}
  \Str(M_{m n} M_{p q}) & = & \eta_{m p} \eta_{n q} - \eta_{m q}
  \eta_{n p} \ , \quad
  \Str(D D) = -1 \ , \quad
  \Str(R R) = -2 \ ,  \nonumber \\
  \Str(P_m K_n) & = & -2 \eta_{m n} \ , \quad
  \Str( Q_\alpha \hat S_\beta) = -8 i \epsilon_{\alpha \beta} \ ,
  \quad
  \Str( \hat Q_\alpha S_\beta) = -8 i \epsilon_{\alpha \beta} \ .
\end{eqnarray}

The $\mathbb{Z}_4$ automorphism invariant subspaces are
\begin{eqnarray}
  \mathcal{H}_0 & = & \{M_{m n}, \ P_m - K_m, \ R\} \ , \nonumber \\
  \mathcal{H}_1 & = & \{Q_\alpha - S_\alpha, \ \hat Q_\alpha + \hat
  S_\alpha\} \ , \nonumber \\
  \mathcal{H}_2 & = & \{P_m + K_m, \ D\} \ , \nonumber \\
  \mathcal{H}_3 & = & \{Q_\alpha + S_\alpha, \ \hat Q_\alpha - \hat
  S_\alpha\} \ .
\end{eqnarray}

\section{The $\mathrm{osp(6|4)}$ algebra in $\mathrm{so}(1,2)\oplus \mathrm{u}(3)$ basis}\label{sec:osp64-algebra}
The $\mathrm{osp}(6|4)$ algebra's commutation relations are given by
\be
[\lambda_{k\dot l},\lambda_{m\dot n}]=2i(\d_{m\dot l}\lambda_{k\dot n}-\d_{k\dot n}\lambda_{m\dot l})
\ee
\be
[\lambda_{k\dot l},R_{m n}]=2i(\d_{m\dot l}R_{k n}-\d_{n\dot l}R_{k m})
\ee
\be
[R_{mn},R_{kl}]=0,\qquad
[R_{mn},R_{\dot k\dot l}]=\frac{i}{2}
(\d_{m\dot k}\lambda_{n\dot l}
-\d_{m\dot l}\lambda_{n\dot k}
-\d_{n\dot k}\lambda_{m\dot l}
+\d_{n\dot l}\lambda_{m\dot k})
\ee
\be
[P_a,P_b]=0,\qquad [K_a,K_b]=0, \qquad
[P_a,K_b]=2\eta_{ab}D-2M_{ab}
\ee
\be
[M_{ab},M_{cd}]=
 \eta_{ac}M_{bd}
+\eta_{bd}M_{ac}
-\eta_{ad}M_{bc}
-\eta_{bc}M_{ad}
\ee
\be
[M_{ab},P_c]=\eta_{ac}P_b-\eta_{bc}P_a, \qquad
[M_{ab},K_c]=\eta_{ac}K_b-\eta_{bc}K_a
\ee
\be
[D,P_a]=P_a, \qquad
[D,K_a]=-K_a, \qquad
[D,M_{ab}]=0
\ee
\be
[D,Q^l_\a]=\half Q^l_\a, \qquad [D,S^l_\a]=-\half S^l_\a
\ee
\be
[P_a,Q^l_\a]=0, \qquad [K_a,S^l_\a]=0
\ee
\be
[P_a,S^l_\a]=-i(\g_a)_{\a}{}^{\b}Q^l_\b, \qquad [K_a,Q^l_\a]=i(\g_a)_{\a}{}^{\b}S^l_\b
\ee
\be
[M_{ab},Q^l_\a]=-\frac{i}{2}(\g_{ab})_{\a}{}^{\b}Q^l_\b, \qquad
[M_{ab},S^l_\a]=-\frac{i}{2}(\g_{ab})_{\a}{}^{\b}S^l_\b
\ee
\be
[R_{kl},Q^{\dot p}_\a]=i(\d^{\dot p l}Q^k_\a-\d^{\dot p k}Q^l_\a)
,\qquad
[R_{kl},S^{\dot p}_\a]=-i(\d^{\dot p l}S^k_\a-\d^{\dot p k}S^l_\a)
\ee
\be
[R_{\dot k\dot l},Q^{p}_\a]=-i(\d^{p\dot  l}Q^{\dot k}_\a-\d^{p \dot k}Q^{\dot l}_\a)
,\qquad
[R_{\dot k\dot l},S^{ p}_\a]=i(\d^{ p \dot l}S^{\dot k}_\a-\d^{p \dot k}S^{\dot l}_\a)
\ee
\be
[\lambda_{k\dot l},Q^{ p}_\a]=2i\d^{p \dot l}Q^k_\a, \qquad
[\lambda_{k\dot l},S^{ p}_\a]=2i\d^{p \dot l}S^k_\a
\ee
\be
[\lambda_{k\dot l},Q^{\dot p}_\a]=-2i\d^{\dot p k}Q^{\dot l}_\a, \qquad
[\lambda_{k\dot l},S^{\dot p}_\a]=-2i\d^{\dot p k}S^{\dot l}_\a
\ee
\be
\{Q^l_\a,Q^k_\b\}=0, \qquad
\{Q^l_\a,Q^{\dot k}_\b\}=-\d^{l\dot k}(\g^a C)_{\a\b}P_a
\ee
\be
\{S^l_\a,S^k_\b\}=0, \qquad
\{S^l_\a,S^{\dot k}_\b\}=-\d^{l\dot k}(\g^a C)_{\a\b}K_a
\ee
\be
\{Q^l_\a,S^k_\b\}=-C_{\a\b}R_{lk}, \qquad
\{Q^{\dot l}_\a,S^{\dot k}_\b\}=-C_{\a\b}R_{\dot l\dot k}
\ee
\be
\{Q^l_\a,S^{\dot k}_\b\}=-i\d^{l\dot k}(C_{\a\b}D+i\half(\g^{ab}C)_{\a\b}M_{ab})+\frac{1}{2}C_{\a\b}\lambda_{l\dot k}
\ee
\be
\{Q^{\dot l}_\a,S^{ k}_\b\}=i\d^{\dot lk}(C_{\a\b}D-i\half(\g^{ab}C)_{\a\b}M_{ab})+\frac{1}{2}C_{\a\b}\lambda_{ k \dot l}
\ee
The indices take the values $k,l=1,2,3$ and the same for the dotted ones --- the $\mathbf{3}$ and $\bar{\mathbf{3}}$ of $\mathrm{u}(3)$, $a,b=0,1,2$ are the $\mathbf{3}$ of $\mathrm{so}(1,2)$ and $\a,\b,..=1,2$ are the $\mathrm{so}(2,1)$ spinors, and $\eta=\textrm{diag}(-,+,+)$. The generators satisfy $R^*_{kl}=R_{\dot k\dot l}$ and $\lambda_{k\dot l}=\lambda^*_{\dot l k}$ and $Q^{\dot l}_\a=(Q^{l}_\a)^*$.
The $(\g_a)_\a{}^\b$ are the Dirac matrices of $\mathrm{so}(1,2)$, and $\g_{ab}=\frac{i}{2}[\g_a,\g_b]$. We raise and lower spinor indices as explained in appendix \ref{sec:psu112^2-algebra}.\\
The bilinear forms are given by
\be\label{bilinear_forms_osp64}
\Str (R_{kl} R_{\dot p\dot q})=-2(\d^{k\dot p}\d^{l\dot q}-\d^{k\dot q}\d^{l\dot p})
\ee
\be
\Str (Q^l_{\a} S^{\dot k}_{\b})=2i\d^{l\dot k}C_{\a\b}
\ee
\be
\Str (P_a K_b)=-2\eta_{ab}
\ee
\be
\Str (D D)=-1
\ee
\be
\Str (M_{ab} M_{cd})=\eta_{ac}\eta_{bd}-\eta_{ad}\eta_{bc}
\ee
The $\mathbb{Z}_4$ automorphism matrix is given by
\begin{equation}
  \Omega = \left(
  \begin{array}{ccc|cc}
    i\s_2 & 0 & 0 & 0 & 0\\
    0 & i\s_2 & 0 & 0 & 0\\
    0 & 0 & i\s_2 & 0 & 0\\ \hline
    0 & 0 & 0 & \s_2 & 0\\
    0 & 0 & 0 & 0 & -\s_2
  \end{array}
  \right) \ .
\end{equation}
The $\mathbb{Z}_4$ invariant subspaces of the algebra are
\begin{eqnarray}
  \mathcal{H}_0 & = & \{ P_a - K_a \ , M_{ab}, \lambda_{l\dot k} \} \ , \nonumber \\
  \mathcal{H}_1 & = & \{ Q^l_{\a}-S^l_{\a},Q^{\dot l}_{\a}-S^{\dot l}_{\a}\} \ , \nonumber \\
  \mathcal{H}_2 & = & \{ P_a + K_a, D,R_{kl},R_{\dot k\dot l} \} \ , \nonumber \\
  \mathcal{H}_3 & = & \{ Q^l_{\a}+S^l_{\a},Q^{\dot l}_{\a}+S^{\dot l}_{\a}\} \ .
\end{eqnarray}

\section{The $\mathrm{psu}(1,1|2)$ algebra} \label{sec:psu112-algebra}

The algebra of the $\mathrm{PSU}(1,1|2)$ group was developed by using
the definition given in \cite{Berkovits:1999zq} (which follows the
definition given in \cite{Kac:1977qb} applied to matrices whose
elements are all Grassmann-even).

The $\mathrm{su}(1,1|2)$ algebra in a basis oriented towards the
$\mathrm{SU}(2)$ R-symmetry is
\begin{eqnarray}
  [D, P] & = & P \ , \quad
  [D, K] = -K \ , \quad
  [P, K] = -2 D \ , \quad
  [R_i, R_j] = -\epsilon_{i j k} R_k \ , \nonumber \\
  {}[D, Q_\alpha] & = & \frac{1}{2} Q_\alpha \ , \quad
  [D, \hat Q_\alpha] = \frac{1}{2} \hat Q_\alpha \ , \quad
  [D, S_\alpha] = -\frac{1}{2} S_\alpha \ , \quad
  [D, \hat S_\alpha] = - \frac{1}{2} \hat S_\alpha \ , \nonumber
  \\
  {}[P, Q_\alpha] & = & [P, \hat Q_\alpha] = 0 \ , \quad
  [P, S_\alpha] = i \hat Q_\alpha \ , \quad
  [P, \hat S_\alpha] = i Q_\alpha \ , \nonumber \\
  {}[K, Q_\alpha] & = & i \hat S_\alpha \ , \quad
  [K, \hat Q_\alpha] = i S_\alpha \ , \quad
  [K, S_\alpha] = [K, \hat S_\alpha] = 0 \ , \nonumber \\
  {}[R_{i j}, Q_\alpha] & = & -\frac{1}{2} {\sigma_{i j \alpha}}^\beta
  Q_\beta \ , \quad
  [R_{i j}, \hat Q_\alpha] = -\frac{1}{2} {\sigma_{i j \alpha}}^\beta
  \hat Q_\beta \ , \nonumber \\
  {}[R_{i j}, S_\alpha] & = & -\frac{1}{2} {\sigma_{i j \alpha}}^\beta
  S_\beta \ , \quad
  [R_{i j}, \hat S_\alpha] = -\frac{1}{2} {\sigma_{i j \alpha}}^\beta
  \hat S_\beta \ , \nonumber \\
  \{Q_\alpha, \hat Q_\beta\} & = & \epsilon_{\alpha \beta} P \ , \quad
  \{S_\alpha, \hat S_\beta\} = \epsilon_{\alpha \beta} K \ , \quad
  \{Q_\alpha, S_\beta\} = -\frac{i}{2} \epsilon_{\beta \gamma}
  \sigma_\alpha^{i j \gamma} R_{i j} - i \epsilon_{\alpha \beta}D
  \ ,\nonumber \\
  \{Q_\alpha, \hat S_\beta\} & = & \frac{i}{2} \epsilon_{\alpha \beta}
  \mathbf{1} \ , \quad
  \{\hat Q_\alpha, S_\beta\} = -\frac{i}{2} \epsilon_{\alpha \beta}
  \mathbf{1} \ , \quad
  \{\hat Q_\alpha, \hat S_\beta\} = \frac{i}{2} \epsilon_{\beta
    \gamma} \sigma_\alpha^{i j \gamma} R_{i j} + i \epsilon_{\alpha
  \beta} D \ , \nonumber \\
  {}
\end{eqnarray}
where $a, b = 1, 2$ are AdS$_2$ indices, $\alpha, \beta, \gamma = 1,
2$ are $\mathrm{SU}(2)$ R-symmetry indices, $R_i$ ($i = 1, \dots, 3$)
are the generators of the $\mathrm{SU}(2)$ R-symmetry and $R_{i j} =
\epsilon_{i j k} R_k$ are defined for convenience. The antisymmetric
tensors are defined such that $\epsilon_{1 2} = 1$, $\epsilon_{1 2 3}
= 1$ and the generators of $\mathrm{SU}(2)$ are ${\sigma_{i j
    \alpha}}^\beta = \frac{1}{2} \left( \sigma_{i \alpha \gamma}
\sigma_j^{\gamma \beta} - \sigma_{j \alpha \gamma} \sigma_i^{\gamma
  \beta} \right)$ with $\sigma_i$ being the Pauli matrices. In order
to get the $\mathrm{psu}(1, 1|2)$ algebra, one has to divide by the
$\mathrm{U}(1)$ generator $\mathbf{1}$. The Grassmann-odd generators
in the algebra above are not Hermitian but are formed as linear
combinations of the original Hermitian matrices with complex
coefficients. These combinations correspond to multiplying the
original Hermitian generators by the complex Killing spinors found by
requiring the supercharges and superconformal transformations to form
Abelian subalgebras.

The non-trivial elements of the Cartan-Killing bilinear form are
\begin{eqnarray}
  \Str (P K) & = & -1 \ , \quad
  \Str (D D) = \frac{1}{2} \ , \quad
  \Str (R_i R_j) = \frac{1}{2} \delta_{i j} \ , \nonumber \\
  \Str (Q_\alpha S_\beta) & = & -i \epsilon_{\alpha \beta} \ , \quad
  \Str (\hat Q_\alpha \hat S_\beta) = i \epsilon_{\alpha \beta}
\end{eqnarray}

The $\mathbb{Z}_4$ automorphism is the same one as given in
\cite{Berkovits:1999zq},
\begin{equation}
  \Omega = \left(
  \begin{array}{cc}
    \sigma_3 & 0 \\
    0 & i \sigma_3
  \end{array}
  \right) \ .
\end{equation}
The $\mathbb{Z}_4$ invariant subspaces of the algebra defined as
\begin{equation}
  \mathcal{H}_k = \left\{ X \in \mathrm{psu}(1, 1|2) | \Omega X
  \Omega^{-1} = i^k X \right\}
\end{equation}
are
\begin{eqnarray}
  \mathcal{H}_0 & = & \{ P + K \ , R_3 \} \ , \nonumber \\
  \mathcal{H}_1 & = & \{ Q_2 + S_2, \hat Q_2 + \hat S_2, Q_1 - S_1,
  \hat Q_1 - \hat S_1 \} \ , \nonumber \\
  \mathcal{H}_2 & = & \{ P - K, D, R_1, R_2 \} \ , \nonumber \\
  \mathcal{H}_3 & = & \{ Q_1 + S_1, \hat Q_1 + \hat S_1, Q_2 - S_2,
  \hat Q_2 - \hat S_2\} \ .
\end{eqnarray}
Another $\mathbb{Z}_4$ automorphism is given by
\begin{equation}
  \Omega = \left(
  \begin{array}{cc}
    \sigma_1 & 0 \\
    0 & i \mathbf{I}
  \end{array}
  \right)
\end{equation}
yields the following invariant subspaces of the algebra
\begin{eqnarray}\label{eq:Z4structure_psu112_AdS2}
  \mathcal{H}_0 & = & \{ D \ , R_i \} \ , \nonumber \\
  \mathcal{H}_1 & = & \{ Q_\a , \hat Q_\a \} \ , \nonumber \\
  \mathcal{H}_2 & = & \{ P \ , K \} \ , \nonumber \\
  \mathcal{H}_3 & = & \{ S_\a , \hat S_\a \} \ .
\end{eqnarray}
This will give us the supercoset space
$\mathrm{PSU}(1,1|2)/(\mathrm{U}(1)\times \mathrm{SU}(2))$ which is isomorphic
to $AdS_2$ with eight supersymmetries.

\section{The $\mathrm{psu}(1,1|2)\oplus \mathrm{psu}(1,1|2)$
  algebra} \label{sec:psu112^2-algebra}
We build the super-algebra by taking the direct sum of two
$\mathrm{psu}(1,1|2)$ algebras (where we take complex combinations of the odd
generators), so the bosonic part is $\mathrm{su}(2) \oplus \mathrm{su}(2)\simeq \mathrm{so}(4)$
and $\mathrm{su}(1,1)\oplus \mathrm{su}(1,1)\simeq \mathrm{so}(2,2)$. The non-vanishing part of
the $\mathrm{psu}(1,1|2)\oplus \mathrm{psu}(1,1|2)$ algebra, involving odd generators, is given by
\be\label{eq:psu(1,1|2)^2Alg}
[D,P_a]=P_a,\quad
[D,K_a]=-K_a,\quad
[D,J_{ab}]=0,\quad
\ee
$$
[P_a,K_a]=2(\eta_{ab}D+J_{ab}),\quad
[P_c,J_{ab}]=\eta_{ac}P_b-\eta_{bc}P_a,\quad
[K_c,J_{ab}]=\eta_{ac}P_b-\eta_{bc}K_a,\quad
$$
$$
[R_\m,R_\n]=-N_{\m\n},\quad
[R_\r,N_{\m\n}]=\d_{\r\m}R_\n-\d_{\r\n}R_\m,\quad
$$
$$
[N_{\r\s},N_{\m\n}]=\d_{\r\m}N_{\s\n}+\d_{\s\n}N_{\r\m}-\d_{\r\n}N_{\s\m}-\d_{\s\m}N_{\r\n}
$$
$$ [P_a,Q^{I}_{\a\ah}]=0,\qquad
         [K_a,S^{I}_{\a\ah}]=0,\qquad [D,Q^{I}_{\a\ah}]=\half
         Q^{I}_{\a\ah},\qquad [D,S^{I}_{\a\ah}]=-\half
         S^{I}_{\a\ah},\qquad
$$
$$
[K_a,Q^{I}_{\a\ah}]=i(\hat \g_a)_\ah{}^{\bh}S^{I}_{\a\bh},\qquad
[P_a,S^{I}_{\a\ah}]=i(\hat \g_a)_\ah{}^{\bh}Q^{I}_{\a\bh},\qquad
$$
$$
[J_{01},Q^{I}_{\a\ah}]=-\frac{i}{2} (\hat \g_{01})_\ah{}^{\bh}Q^{I}_{\a\bh},\qquad
[J_{01},S^{I}_{\a\ah}]=-\frac{i}{2} (\hat \g_{01})_\ah{}^{\bh}S^{I}_{\a\bh},\qquad
$$
$$
[R_\m,S^{I}_{\a\ah}]=\frac{i}{2}(i\hat \g_{01})_\ah{}^{\bh}(\g_\m)_{\a}{}^{\b}S^{I}_{\b\bh},\qquad
[R_\m,Q^{I}_{\a\ah}]=-\frac{i}{2}(i\hat \g_{01})_\ah{}^{\bh}(\g_\m)_{\a}{}^{\b}Q^{I}_{\b\bh},\qquad
$$
$$
[N_{\m\n},S^{I}_{\a\ah}]=\frac{i}{2}\hat C_\ah{}^{\bh}(\g_{\m\n})_{\a}{}^{\b}S^{I}_{\b\bh},\qquad
[N_{\m\n},Q^{I}_{\a\ah}]=\frac{i}{2}\hat C_\ah{}^{\bh}(\g_{\m\n})_{\a}{}^{\b}Q^{I}_{\b\bh},\qquad
$$
$$
\{S^{I}_{\a\ah},S^{J}_{\b\bh}\}=\frac{i}{2}\e^{IJ}C_{\a\b}(\hat \g^a \hat C)_{\ah\bh}K_a,\qquad
\{Q^{I}_{\a\ah},Q^{J}_{\b\bh}\}=\frac{i}{2}\e^{IJ}C_{\a\b}(\hat \g^a \hat C)_{\ah\bh}P_a,\qquad
$$
$$
\{S^{I}_{\a\ah},Q^{J}_{\b\bh}\}
=\frac{i}{2}\e^{IJ}(C_{\a\b}(i\hat C_{\ah\bh}D-\frac{1}{2}(\hat C\hat \g^{ab})_{\ah\bh}J_{ab})
$$
$$
+\frac{1}{2}\hat C_{\ah\bh}N_{\m\n}(\g^{\m\n}C)_{\a\b}-i(\hat \g_{01}\hat C)_{\ah\bh}R_\m(\g^\m C)_{\a\b})
$$
where the indices go as $I=1,2$, $\ah=1,2$, $\a=1,2$,$a=1,2$, $\m=1,2,3$ and $C_{\a\b}=\e_{\a\b}$, $\hat C_{\ah\bh}=\e_{\ah\bh}$.
The gamma matrices are defined by $(\hat \g_a)_\ah{}^\bh=\{i\s_2,-\s^1\}$, $(\hat \g_{01})_\ah{}^\bh=\frac{i}{2}[\hat \g_0,\hat \g_1]_\ah{}^\bh=\{-i\s^3\}$, $(\g_\m)_\a{}^\b=\{\s_1,\s_3,\s_2\}$, $(\g_{\m\n})_\a{}^\b=\frac{i}{2}[\g_\m,\g_\n]_\a{}^\b=\{\s_2,\s_1,\s_3\}$, where $\s$ are the Pauli matrices, and $\eta_{ab}=\mathrm{diag}(+,-)$.
We raise and lower spinor indices using $\psi_\a=\psi^\b\e_{\b\a}$, $\psi^\a=\e^{\a\b}\psi_\b$, where $\e_{12}=-\e_{21}=\e^{12}=-\e^{21}=1$ and the same for hatted objects.
The bilinear Cartan-Killing forms can be rescaled to give
\be
\Str (P_a K_b)=\eta_{ab},\qquad
\Str (DD)=1,\quad
\Str (J_{ab}J_{cd})=\d_{ac}\d_{bd}-\d_{ad}\d_{bc}
\ee
$$
\Str (R_\m R_\n)=\d_{\m\n}, \qquad
\Str (N_{\m\n}N_{\r\s})=\d_{\m\r}\d_{\n\s}-\d_{\m\s}\d_{\n\r}
$$
$$
\Str (Q^{I}_{\a\ah} S^{J}_{\b\bh})=\e^{IJ}\e_{\a\b}\e_{\ah\bh}
$$
The super-algebra has $\mathbb{Z}_4$ grading structure using the automorphism matrix
\be
\O=\left( \begin{array}{cc}
\begin{array}{cc}
0\quad & 0 \\
0\quad & 0 \\
\end{array}
 & \begin{array}{cc}
\s_3 & 0 \\
0 & 1_{2\times 2} \\
\end{array} \\
\begin{array}{cc}
-i\s_3 & 0 \\
0 & i1_{2\times 2} \\
\end{array} &
\begin{array}{cc}
0\quad & 0 \\
0\quad & 0 \\
\end{array}
 \\
\end{array} \right)
\ee
so
\be\label{eq:ads3S3Z4struct}
\mathcal{H}_0=\{J_{01},N_{\m\n},P_0-K_0,P_1-K_1\}
\ee
\be
\mathcal{H}_1=\{S^{I}_{\a\ah}+a^{IJ}Q^{J}_{\a\ah}\}
\ee
\be
\mathcal{H}_2=\{R_{\m},D,P_0+K_0,P_1+K_1\}
\ee
\be
\mathcal{H}_3=\{S^{I}_{\a\ah}-a^{IJ}Q^{J}_{\a\ah}\}
\ee
where $a^{IJ}=\s_1$. So we can get the supercoset space $\mathrm{PSU}(1,1|2)^2/(\mathrm{SU}(1,1)\times \mathrm{SU}(2))$, whose bosonic part is $AdS_3\times S^3$.
We also have another $\mathbb{Z}_4$ grading structure with the same number of supersymmetry generators
\be
\mathcal{H}_0=\{D,J_{01},N_{\m\n},R_{\m}\}
\ee
\be
\mathcal{H}_1=\{Q^{I}_{\a\ah}\}
\ee
\be
\mathcal{H}_2=\{P_a,K_a\}
\ee
\be
\mathcal{H}_3=\{S^{I}_{\a\ah}\}
\ee
which is the four-dimensional space $BDI(2,2)\simeq AdS_2\times AdS_2$, given by the supercoset $\mathrm{PSU}(1,1|2)^2/(\mathrm{SO}(2)^2\times \mathrm{SO}(4))$.

\bibliography{postdoc}
\bibliographystyle{JHEP}
\end{document}